\documentclass[aps,pra,twocolumn,groupedaddress,floatfix]{revtex4-2}
\usepackage{graphicx}
\usepackage{amsmath}
\usepackage{amssymb}
\usepackage{braket}
\usepackage{eqnarray}
\usepackage{dsfont}
\usepackage{multirow}
\usepackage[T1]{fontenc}

\newcommand\sqnorm{\frac{1}{\sqrt{2}}}
\newcommand\dnorm{\frac{1}{\sqrt{D}}}
\newcommand\psucc{p_\text{succ}}
\newcommand\plink{p_\text{link}}
\newcommand\tcoh{t_\text{coh}}
\newcommand{\ppauli}[1]{\mathbf{P}_\text{Pauli}^{\text{#1}}}

\begin{document}

\title{Quantum repeaters based on stationary and flying Gottesman-Kitaev-Preskill qudits}
\author{Stefan H{\"a}ussler and Peter van Loock}
\affiliation{Institute of Physics, Johannes Gutenberg-Universit{\"a}t Mainz, Staudingerweg 7, 55128 Mainz, Germany}
\date{\today}

\begin{abstract}
There are various approaches to long-range quantum communication
based on conceptually different forms of quantum repeaters.
Here we explore a quantum repeater scheme that employs 
quantum error correction (QEC) both on the flying (light) qubits and on the
stationary (matter) qubits. 
The idea is to combine the benefits of encoded one-way and two-way schemes
where effective channel transmission and loss scaling are enhanced by means of 
photon loss codes and encoded quantum memories, respectively,
while sacrificing some of their advantages such as high clock rates, 
independent of classical communication times (one-way), and 
potentially large segment lengths (two-way).
More specifically, we illustrate, propose, and analyze such a quantum 
repeater using the bosonic Gottesman-Kitaev-Preskill (GKP) code which naturally 
enables encoding and QEC of qudits, protecting them against transmission 
and memory loss, the latter, for instance, occuring on collective spin modes
of atomic ensembles. 
While the encoded one-way and two-way schemes on their own
either require very high repeater link coupling efficiencies and GKP squeezing
or allow for experimentally more feasible, small values of these parameters, respectively,
we find that there are intermediate parameter regimes where the combined
repeater protocol is superior.                 
\end{abstract}

\maketitle

\section{Introduction}
Quantum repeaters are a promising solution to overcome the challenge posed to the creation of remote entanglement, constituting a crucial resource for applications such as quantum key distribution (QKD) \cite{BB84} or distributed quantum computing \cite{eco}, by the exponentially decreasing transmissivity of long-distance optical fiber channels.
Typically, the effect of transmission of a quantum state through an optical fiber of length $L$ is modeled as a bosonic loss channel
\begin{equation}
\label{app::eq::losschannel}
\mathcal{L}_\eta[\rho] = \text{tr}_E \left(B(\eta)\rho\otimes\ket{0}_E\bra{0}B^\dagger(\eta)\right),
\end{equation}
where $B$ denotes the unitary operator corresponding to a beam splitter, and the transmissivity is related to the total distance by the exponential decay $\eta \propto e^{-L/L_\text{att}}$ governed by the fiber's characteristic attenuation length $L_\text{att} = 22$km,
which also serves as an upper bound on the transmission rates \cite{Takeoka2014}.
As shown by Pirandola et al. in Ref.~\cite{plob}, the secret-key distribution capacity of such a channel reads 
\begin{equation}
C[\mathcal{L}] = -\log_2(1 - \eta)\approx 1.44\, \eta ,
\end{equation} 
wherein the approximation is valid for small $\eta$.
This relation, commonly referred to as the PLOB bound, restricts the rate at which distribution of entanglement across large distances can be performed by direct transmission; however, it can be overcome when turning the transmission line into a so-called quantum repeater chain by introducing a number of intermediate stations along the way and thus partitioning the total distance into several shorter segments. Then the (repeater-assisted) quantum capacity of the total link is given by the minimum of the individual segments' capacities \cite{repeaterassisted}; in the optimal case of evenly spaced stations separated by a distance $L_0$, it therefore reads 
\begin{equation}
C^\text{(rep. ass.)}[\mathcal{L}] = -\log_2(1 - \eta_0),
\end{equation} 
with $\eta_0 \propto e^{-L_0/L_\text{att}}$ only scaling with the length of a single segment. 
Another thing to note is that the capacity can exceed one bit per channel use when $\eta_0 > 1/2$, which corresponds to an upper bound on the segment length of $L_0 < 15$km. In such cases, it may be possible to enhance QKD rates through the use of higher-dimensional states, referred to as qudits, instead of the usual two-dimensional qubits, as they can represent more than one bit of classical information. 

Since the first quantum repeater proposal \cite{firstrepeater} in 1998, a multitude of other proposals have followed. While they all share the basic idea of partitioning the total distance into shorter segments, the way that transmission loss and operational errors are addressed varies greatly; however, three main paradigms, commonly referred to as repeater generations, have been established to categorize quantum repeaters \cite{generations}.
The first and second generations both employ quantum memories to store successfully distributed entangled states created during probabilistic entanglement distribution attempts in each segment, and connect states of neighboring segments via entanglement swapping in order to gradually obtain remote entanglement. However, they differ in the way that errors caused by memory loss or imperfections are addressed, with the first generation using entanglement distillation and the second generation using quantum error correction (QEC) \cite{firstrepeater, secondrepeater}.
In contrast, the third generation dispenses with quantum memories in favor of an all-optical setup where encoded logical qubits (or qudits) are sent through the fiber links instead of single-photon states to provide some form of protection from transmission loss \cite{thirdgen_surfacecode, Munro2012, Azuma2015, Muralidharan2014, Ewert2016, Ewert2017, Lee2019}. 
The lack of memories makes it necessary that all distribution attempts succeed simultaneously; however, this disadvantage is offset by a significantly higher repetition rate enabled by the all-optical design as well as the absence of two-way classical communication.
In principle, the third generation can achieve the highest rates (per second) under favorable conditions; however, it also imposes the highest demands on experimental parameters such as fiber coupling efficiency and gate fidelities, and is thus challenging to implement in practice. 

Looking at the QEC-based repeater generations, one notices how QEC is used in two distinct ways to benefit the communication protocol, either protecting stationary states on the memories, or flying states being transmitted through the fiber. 
In this work, we introduce what could be referred to as a new repeater generation that combines both approaches. More precisely, we propose a memory-based repeater with third-generation-style encoded entanglement distribution in each segment and teleportation-based QEC during entanglement swapping on the memories, in the hope of improving the distribution success probability compared to the second generation and lessening the hardware demands compared to the third generation. 
However, based upon memories and two-way classical communication, the combined repeater scheme no longer exhibits the high clock rates of a one-way error-corrected third-generation scheme. At the same time, relying upon photonic quantum error correction codes employed in the optical fiber channel, the combined repeater requires sufficiently frequently placed repeater stations for transmission loss correction, unlike the second-generation schemes that potentially allow for relatively large segment lengths. 
Nonetheless, to assess the usefulness of this new repeater generation, we will compare its performance with that of the established QEC-based generations, i.e. the second and third generations, for both qubits as well as higher-dimensional logical states. As the error correction code, we choose the Gottesman-Kitaev-Preskill (GKP) code \cite{gkp}, which has formed the basis for several previous repeater proposals \cite{häussler_vanloock, thirdgen_gkp, alloptical, Rozpedek2021, Rozpedek2023}.
\\

The paper is structured as follows: In Secs.~\ref{sec::ESdis} to \ref{sec::rateanalysis}, we review (and generalize beyond the special case of qubits) the discussion of the building-blocks and rate analysis of a GKP-based second-generation quantum repeater from Ref.~\cite{häussler_vanloock}. In Sec.~\ref{sec::thirdgen}, we give a short review of third-generation GKP repeaters as previously discussed in more detail in Refs.~\cite{thirdgen_gkp} and \cite{alloptical}. In Sec.~\ref{sec::fourthgen}, we propose a new fourth repeater generation as a synthesis of the second and third generations, and discuss its implementation based on the GKP code. Additionally, we explain how to perform rate analysis for this new generation. 
Finally in Sec.~\ref{sec::comparison}, we compare the secret key rates that can be achieved in a QKD context by the established generations with those of our fourth generation, investigating points such as squeezing demands or the effect of imperfect fiber-coupling efficiency.

\section{GKP quantum repeaters}
\subsection{Second-generation GKP repeaters}
\label{sec::secondgen}
In this section, we will review the functioning principle and the rate analysis of a second-generation quantum repeater based on the bosonic GKP encoding. In contrast to Ref.~\cite{häussler_vanloock}, where a similar situation was discussed specifically for qubits, the dimension of the logical state space may now exceed $D = 2$. 
Second-generation GKP quantum repeater here means that the flying qudits are single-photon-based, only allowing for transmission loss detection, and the stationary qudits are GKP qudits, enabling memory loss correction. 
\begin{figure}
\includegraphics[width=0.45\textwidth]{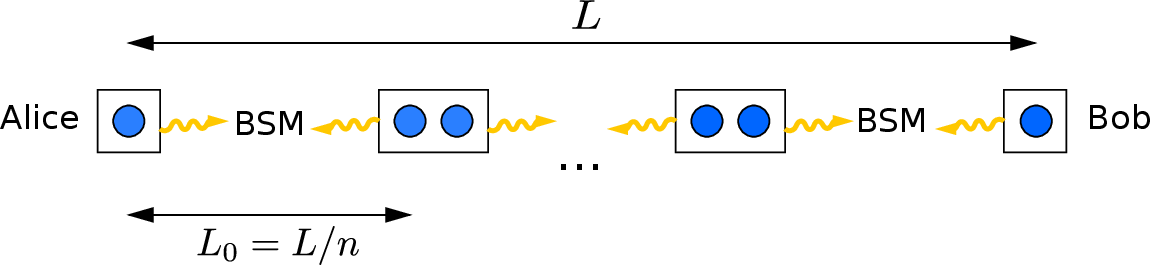}
\caption{(Color online) Schematic illustration of a memory-based quantum repeater: the total distance $L$ is partitioned into $n$ segments of length $L_0$ by inserting intermediate stations. Every station contains two quantum memories to enable entanglement swapping between neighboring segments. In this paper, we consider quantum repeaters where these stationary states are logical qudits protected against memory (storage and gate) errors by a suitable quantum error correction code, specifically the GKP code. The encoded entanglement distribution is achieved by locally entangling optical qudits with stationary GKP qudits, storing the latter, and sending the former towards the middle of each segment, where a Bell state measurement (BSM) is performed. }
\label{fig::setup}
\end{figure}

\subsubsection{Entanglement distribution}
\label{sec::ESdis}

The most basic step in any memory-based repeater protocol is the distribution of entanglement in the form of one of the $D^2$ maximally entangled Bell states
\begin{equation}
\label{eq::bellstates}
    \ket{\phi^{ab}} = \dnorm \sum_{k = 0}^{D-1}e^{ikb}\ket{k}\otimes\ket{k+a \mod D},
\end{equation}
with $0\leq a, b\leq D-1$,
between the memories forming the ends of each segment. In this paper, we consider memories that can support a bosonic mode Hilbert space, allowing for the use of bosonic QEC codes, in particular the GKP code, to protect against memory loss. A promising candidate for such a memory exists in the form of atomic spin ensembles \cite{hammerer}. GKP-like states can be defined both in the case of relatively few atoms \cite{ensgkp} as well as via the Holstein-Primakoff approximation in the case of large ensembles \cite{hp}.   
To achieve entanglement across the length of each segment, entangled states between the memory modes and a photonic system must first be created locally at the repeater stations. We assume this, as well as the creation of GKP states in general, to be a deterministic process. The photonic components, typically taking the form of multi-rail encoded single photon states, are then sent towards the segment's midpoint, where a Bell state measurement is performed in order to obtain entanglement between the states in the memories.  
In the qubit case, the Bell-state measurement can be performed by means of linear optics with a success probability of $1/2$ without additional resources \cite{Calsamiglia2001}, which can be increased arbitrarily close to $1$ when supplemented with auxiliary photons \cite{Knill2001, Grice2011, Ewert_bsm, Bayerbach2023, boostedfusiongates}. Contrarily, auxiliary photons are a necessary resource for the Bell state measurement of qudits with dimension greater than $2$, as it has been shown that without them, no measurement scheme restricted to linear optics can implement a POVM element with a Schmidt number exceeding $2$ \cite{Calsamiglia2002}. This also implies that a perfect discrimination of qudit Bell states is impossible even with auxiliary photons \cite{Carollo}; however, there are proposals for measurement schemes operating at a non-unit success probability similar to the case of qubits. In particular, the success probability of the scheme introduced in Ref.~\cite{Dusek} can theoretically be brought arbitrarily close to $1$, albeit at the cost of requiring an increasing number of auxiliary photons in complex entangled states. Recently, another proposal has been put forward that is applicable to qudits of even dimensions \cite{Bharos}, and can achieve a success probability of $\psucc^\text{(BSM)} = 2/{D^2}$ using only $D - 2$ auxiliary photons in an entangled state of Schmidt number $D/2$.

Besides failure being caused by the possibly limited efficiency of the Bell state measurement at the midpoint, the entanglement distribution process in one segment can fail if photons sent from the endpoints are lost before reaching the middle. Two main effects can be distinguished that contribute to this: firstly, absorption and scattering in the fiber, and secondly, losses occurring when coupling light into  and out of the fiber. The former is quantified in the form of a factor decreasing exponentially with segment length $L_0$, while the latter is taken into account by a constant $\plink$. Note that in this paper, we define $\plink$ to represent the efficiency of one photon coupling into and later out of the fiber once, such that the success probability in one segment scales with $\plink^2$ due to the segment being composed of two separate fibers, one from the right and left endpoint to the midpoint, respectively.

Thus, entanglement distribution in one segment is successful with probability 
\begin{equation}
\label{eq::psucc2}
p = \psucc^\text{(BSM)}\plink^2\exp\left(-\frac{L_0}{L_\text{att}}\right),
\end{equation} 
where $L_\text{att}$ denotes the attenuation length of the fiber, typically around 22km at telecom wavelengths, 
and the number of attempts until success is distributed geometrically with parameter $p$.

\subsubsection{GKP QEC and entanglement swapping}
\label{sec::GKP}

Since entanglement distribution is probabilistic, it may occur that a given segment achieves successful distribution before any of its neighbors. In this case, states are stored in this segment's memories until a neighboring segment heralds distribution success and entanglement swapping can be performed by means of a logical Bell-state measurement at the repeater station connecting the ``old'' and the ``new'' segment.
A big advantage of the GKP code consists in the possibility of performing deterministic logical Bell-state measurements via homodyne detection even in the qudit case. Instead of reducing the success probability, noise accumulating during the waiting period due to the limited memory coherence time $\tcoh$ manifests as logical Pauli errors after the entanglement swapping.  
\\

The $D$-dimensional GKP code \cite{gkp} is the subspace of the bosonic mode Hilbert space spanned by the idealized basis states
\begin{equation}
\label{eq::GKPdef}
\ket{\overline{j}} = \sum_{k \in \mathds{Z}}\Ket{(kD + j)\sqrt{\frac{2\pi}{D}}}_q,
\end{equation}
where $0 \leq j \leq D-1$ and kets with index $q$ denote $q$-quadrature eigenstates.
Being composed of displaced infinitely squeezed states, these states cannot be realized exactly in practice, but only be approximated by replacing delta-peaks in phase space with narrow Gaussians and truncating the infinite sum on both sides. In theoretical descriptions it is common to omit the truncation and instead modulate the sum by a wide Gaussian envelope; however, here we will make use of the simpler Gaussian noise approximation to account for finite squeezing effects: we define realistic GKP states as resulting from the action of a Gaussian displacement channel on the idealized states:
\begin{equation}
\left(\ket{\overline{j}}\bra{\overline{j}}\right)_\text{real} = \mathcal{E}_{\delta^2}\left[\left(\ket{\overline{j}}\bra{\overline{j}}\right)_\text{ideal}\right],
\end{equation}
with the Gaussian displacement channel with variance $\delta^2$ defined as
\begin{equation}
\label{eq::shiftchannel}
\mathcal{E}_{\delta^2}[\rho] = \frac{1}{\pi\delta^2}\int_\mathds{C} d^2\alpha\, \exp\left(-\frac{|\alpha|^2}{\delta^2}\right)D(\alpha)\rho D^\dagger(\alpha).
\end{equation}
Instead of referring directly to the variance, it is more common to use a squeezing parameter $s$ expressed in units of dB and defined as
\begin{equation}
    s = -10\log_{10}(2\delta^2).
\end{equation}

The GKP code is specifically designed to protect against random Gaussian displacements in phase space, however, since the dominant noise channel in our repeater, both in the memories and later for the fourth generation during the transmission through the fiber, is bosonic loss, an additional step is required to convert the bosonic loss into random Gaussian shifts. Such a conversion can be achieved using amplification \cite{channels}, either in the form of physical preamlification, or via a technique referred to as ``CC-amplification'' \cite{alloptical}, which involves rescaling of measurement results on a classical computer. Important for the rate analysis of GKP-based quantum repeaters is the average variance resulting from the combination of loss and amplification, as this determines the probability of logical Pauli errors resulting from entanglement swapping. 
In App.~\ref{app::sec::amplification_wait} we discuss the amplification strategies in more detail and derive 
\begin{eqnarray}
\label{eq::varwait}
\mathds{E}(\sigma_\text{add}^2) &= \min&\left[\frac{p^2}{1-q^2}\left(\frac{1-e^{-\alpha}}{e^{-\alpha}} + \frac{2e^{2\alpha}q}{1-qe^{\alpha}} - \frac{2q}{1-q}\right)\right.,\nonumber\\
 & & \left.\vphantom{\frac12}(T + 2)(1 - e^{-\alpha})\right]
\end{eqnarray}
as the expression for the average variance added by memory loss and amplification.
Therein, $q = 1 - p$ is a shorthand for the failure probability of entanglement distribution, $T = 2q / (1 - q^2)$ is the average number of timesteps of duration $\tau_0$ that a memory spends waiting for the adjoining segment to finish distribution, and $\alpha = \tau_0 / \tcoh$ denotes the effective inverse coherence time. The necessity for heralding distribution success via classical communication imposes the lower bound $\tau_0 \geq L_0/c$ on the length of timesteps. This will be discussed in more detail in Sec.~\ref{sec::comparison}.
The total average variance influencing the Bell state measurement including the finite squeezing variance $\delta^2$ of both modes involved in the swapping process is thus given by
\begin{equation}
\sigma^2 = 2\delta^2 + \mathds{E}(\sigma_\text{add}^2).
\end{equation}

\begin{figure}
\includegraphics[width=0.45\textwidth]{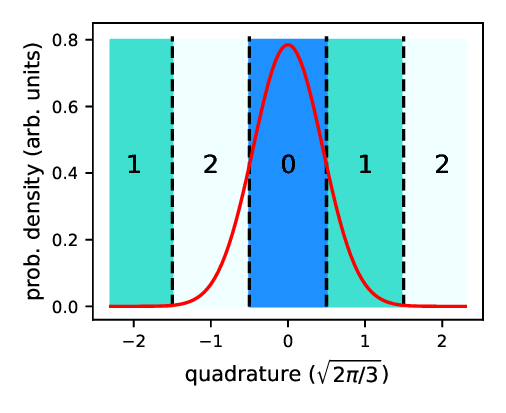}
\caption{(Color online) Pauli errors from GKP QEC for $D=3$. The Gaussian curve represents the probability distribution of the ``true'' shift, the numbers $0$ to $2$ label the Pauli operator power acting on the resulting state if the shift lies within the corresponding colored error region.}
\label{fig::errorregions}
\end{figure}

In our scheme, teleportation-based QEC \cite{knillstyle} and entanglement swapping are performed simultaneously in one step using the newly formed Bell state in the ``new'' segment as the resource to teleport the adjoining component of the ``old'' segment into the remote component of the ``new'' segment. To this end, the linear combinations $q_A - q_B$ and $p_A + p_B$ of the two adjacent modes' quadratures need to be measured, which can be physically realized for atomic ensemble memories by repeated interaction with suitably polarized electromagnetic fields and subsequent homodyne detection applied to these fields \cite{häussler_vanloock}. The measurement results are interpreted by choosing the closest element of the set $\{k\sqrt{2\pi / D}|k\in\mathds{Z}\}$, dividing by $\sqrt{2\pi / D}$  and taking the result modulo $D$. The numbers obtained in this fashion from the position and momentum linear combination correspond to the parameters $a$ and $b$ characterizing the qudit Bell states $\ket{\phi^{ab}}$ and thus reveal exactly the information required in a Bell state measurement. \\

In a $D$-dimensional space, the Pauli operators can be generalized by defining 
\begin{subequations}
\begin{eqnarray}
X\ket{j} &= &\ket{j+1 \mod D}\\
Z\ket{j} &= &e^{2\pi i j}\ket{j}
\end{eqnarray}
\end{subequations}
as their action on qudit basis states. 
The reinterpretation of measurement results as the closest multiple of $\sqrt{2\pi / D}$ will introduce the $r$-th power of a logical Pauli error
whenever the sum of the ``true'' shifts affecting the memory modes lies within any of the intervals from the set $\{[(kD + r - 1/2)\sqrt{2\pi / D}, (kD + r + 1/2)\sqrt{2\pi / D}]|k\in\mathds{Z}\}$. This situation is illustrated for $D=3$ in Fig.~\ref{fig::errorregions}
In particular, the position quadrature measurement will give rise to powers of $X$ and the momentum quadrature measurement to powers of $Z$; however, since we assume the Gaussian displacement to be symmetric in phase space, the error probabilities will be identical for both quadratures and thus no distinction needs be made in the following.
The probabilities for the post-swapping state carrying the $r$-th power of a Pauli operator can be collected into a zero-indexed list of length $D$ whose components may be calculated by integrating a Gaussian with variance $\sigma^2$ over the appropriate intervals :
\begin{equation}
\label{eq::paulierror}
\left(\ppauli{wait}\right)_r = \sum_{k \in\mathds{Z}} \int_{(kD + r - 1/2)\sqrt{2\pi/D}}^{(kD + r + 1/2)\sqrt{2\pi/D}}\, dx \frac{1}{\sqrt{2\pi\sigma^2}}\exp\left(\frac{x^2}{2\sigma^2}\right).
\end{equation}

\subsubsection{QKD secret key rates}
\label{sec::rateanalysis}
To quantify the performance of our repeater schemes, we use the secret key rate (SKR) $S$, defined as a product of the raw rate $R$ and the secret key fraction (SKF) $r$, as the figure of merit. It describes how many bits of a secret key can be generated per timestep by the repeater in a QKD application like BB84 \cite{BB84}.
The raw rate for a memory-based repeater scheme can be calculated as the inverse of the average number of timesteps $K_n$ required for the total process in an $n$-segment repeater if deterministic entanglement swapping is performed as soon as possible, i.e. $R = 1 / \mathds{E}(K_n)$. The explicit expression 
\begin{equation}
\label{eq::inverse_raw}
\mathds{E}(K_n) = \sum_{i=1}^n (-1)^{i+1}\begin{pmatrix}n\\i\end{pmatrix}\frac{1}{1-q^i },
\end{equation}  
where $q = 1 - p$ once again represents the entanglement distribution failure probability in each segment,
is derived and employed in Refs.~\cite{nadja_rateanalysis, waitingtime_pra, rateanalysis}.

Somewhat more interesting than the raw rate is the SKR's second component, the secret key fraction, as it often proves the factor limiting a repeater scheme's tolerance against imperfect values of parameters like $\plink$, and hinders the extension towards more favorable operating characteristics such as greater total distance. 
The central quantity for the definition of the SKF is the so-called quantum bit error rate (QBER), which we denote by $\boldsymbol{\epsilon}$. Similarly to $\ppauli{wait}$ it is a zero-indexed list of length $D$ whose $r$-th entry represents the probability of the finally distributed state between the remote parties Alice and Bob differing from the expected state by the $r$-th power of a Pauli operator. 
The QBER is obtained from $\ppauli{wait}$ by $n-1$-fold circular convolution:
\begin{equation}
\label{eq::qber_second}
    \boldsymbol{\epsilon} = \left(\ppauli{wait}\right)^{\circledast(n-1)},
\end{equation}
an operation defined on finite lists as
\begin{equation}
    (\mathbf{a} \circledast\mathbf{b})_r = \sum_{i = 0}^{D-1} a_i b_{r - i \mod D} .
\end{equation}
Important for the SKF is the QBER's Shannon entropy. A large Shannon entropy corresponds to a low SKF, as high uncertainty about the occurrence of errors limits the usefulness of the final state for QKD applications. The SKF is accordingly defined as  
\begin{equation}
\label{eq::skf}
    r = \log_2(D) - 2H(\boldsymbol{\epsilon}),
\end{equation}
whenever this expression is positive, and zero otherwise. 
Therein, $H$ denotes the Shannon entropy
\begin{equation}
\label{eq::entropy}
    H(\mathbf{x}) = -\sum_{i = 0}^{D-1} x_i \log_2(x_i) 
\end{equation}
of a probability vector and the factor $2$ accounts for the fact that Pauli-$X$ and $Z$ errors occur with equal probability.
Note that in the case of $D > 2$, the SKF may take values greater than $1$, reflecting the fact that a qudit can carry more than a single classical bit of information.

\subsection{Third-generation GKP repeaters}
\label{sec::thirdgen}
In this section, we discuss elements of third-generation quantum repeaters based on GKP qudits. This means there are no stationary qudits at all and the flying optical qudits are GKP qudits that can be protected to some extent against photon transmission loss in the fiber channel. 

\subsubsection{GKP QEC of transmission loss}
Repeaters of the third generation follow a very different approach, dispensing with memories entirely and applying QEC to counteract transmission loss during the entanglement distribution process \cite{alloptical, thirdgen_gkp}. Consequently, the entangled states initially created are GKP-encoded logical Bell states between two optical modes traveling in opposite directions from the repeater stations. 
During transmission, the optical GKP states are subject to a bosonic loss channel with transmissivity 
\begin{equation}
    \eta_\text{dist} = \plink \exp\left(-\frac{L_0}{2L_\text{att}}\right),
\end{equation}
which, combined with the appropriate amplification detailed in App.~\ref{app::sec::amplification_dist}, results in an added variance of 
\begin{equation}
\label{eq::vardist}
    \chi^2_\text{add} = 
    \begin{cases}  2(1 - \eta_\text{dist})& \eta_\text{thresh} \leq \eta_\text{dist} \leq 1/2\\
        (1 - \eta_\text{dist})/\eta_\text{dist} & \text{else}
    \end{cases}
\end{equation}
affecting the Bell state measurement. 
Here, $\eta_\text{thresh}$ is a parameter related to the limit of practically achievable preamplification strength. 

\begin{figure}
\includegraphics[width=0.45\textwidth]{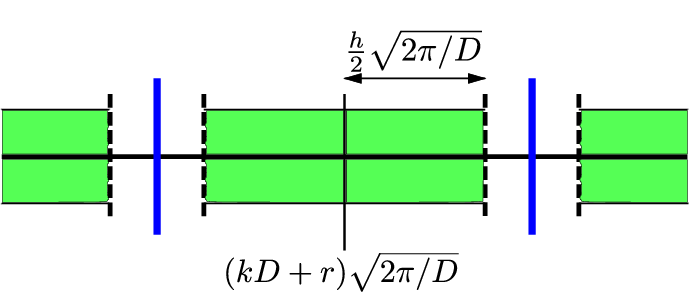}
\caption{(Color online) Principle of HRM \cite{alloptical}: measurement outcomes are only accepted if they fall within the acceptance regions (shaded green) parameterized by $h$. The solid blue vertical lines mark the boundaries between different error regions. }
\label{fig::hrm}
\end{figure}

Upon arrival at a segment's midpoint, a logical BSM is performed on the modes arriving from both ends by interfering them at a balanced beamsplitter followed by homodyne detection on the outputs.  
Contrary to the second generation, the distribution process can in principle be made deterministic due to the removal of the limited efficiency linear-optics Bell measurement and the conversion of losses to logical Pauli errors by the GKP QEC. However, as proposed in Ref.~\cite{alloptical}, it may prove advantageous to forgo deterministic entanglement distribution in favor of reduced Pauli error probabilities.
The idea behind the so called ``highly reliable measurement'' (HRM) \cite{alloptical} is to reject homodyne results that lie too close to the boundaries between different error regions based on the observation that the probability of causing a Pauli error under the condition of measuring any particular value is greatest near these interval boundaries. More formally, we define a parameter $h\in (0, 1)$ and declare acceptance regions of width $h\sqrt{2\pi / D}$ centered around integer multiples of $\sqrt{2\pi / D}$, as illustrated in Fig.~\ref{fig::hrm}. Whenever the homodyne results from both quadratures fall within any of the acceptance regions, the entanglement distribution is considered successful, whereas otherwise it is considered failed and a new attempt will be made in the following timestep.
Consequently, the probability of entanglement distribution success reads
\begin{equation}
\label{eq::psucc4}
p = \left[\sum_{k \in\mathds{Z}} \int_{(k - h/2)\sqrt{2\pi/D}}^{(k + h/2)\sqrt{2\pi/D}}\, dx \frac{1}{\sqrt{2\pi\chi^2}}\exp\left(\frac{x^2}{2\chi^2}\right)\right]^2,
\end{equation}
with $\chi^2 = 2\delta^2 + \chi^2_\text{add}$.

\subsubsection{Third-generation secret key rates}
The lack of memories changes the raw rate of a third generation repeater from the complicated relation in Eq~(\ref{eq::inverse_raw}) to the simple exponential
\begin{equation}
    R = p^n,
\end{equation}
reflecting the fact that entanglement spanning the distance between Alice and Bob is achieved only if distribution succeeds in all segments simultaneously.
In general, memoryless schemes also differ from memory-based ones in their maximum repetition rate; this becomes relevant when considering key rates per unit of time instead of per channel use and will be discussed in more detail in Sec.~\ref{sec::comparison}.

Calculating the SKF proceeds along similar lines as before,
with the vector of Pauli error probabilities, now arising from the entanglement distribution and carrying dependence on the HRM acceptance parameter $h$ within the integral boundaries, given by
\begin{equation}
\label{eq::ppaulidist}
\left(\ppauli{dist}\right)_r = \frac{\sum_{k \in\mathds{Z}} \int_{(kD + r - h/2)\sqrt{2\pi/D}}^{(kD + r + h/2)\sqrt{2\pi/D}}\, dx \frac{1}{\sqrt{2\pi\chi^2}}\exp\left(\frac{x^2}{2\chi^2}\right)}{\sqrt{p}}.
\end{equation} 
The factor of $1/\sqrt{p}$ in the definition of $\ppauli{dist}$ accounts for the fact that purely integrating over all the $r$-th power error regions yields the joint probability of accepting the homodyne result and obtaining an $r$-th power Pauli error, instead of the conditional probability of a Pauli error occurring under the condition of HRM acceptance, which is the quantity of interest for our rate analysis.  
Similarly to the second generation, the QBER can be obtained by repeated circular convolution of the Pauli probability vector with itself:
\begin{equation}
\label{eq::qber_third}
    \boldsymbol{\epsilon} = \left(\ppauli{dist}\right)^{\circledast n}.
\end{equation}
Note that since Pauli errors in a third generation repeater arise from the entanglement distribution within each segment instead of entanglement swapping between segments, the exponent now reads $n$ rather than $n-1$ as in Eq.~(\ref{eq::qber_second}).
Finally, the SKF is found by inserting Eq.~(\ref{eq::qber_third}) into Eq.~(\ref{eq::skf}).

\subsection{The fourth repeater generation}
\label{sec::fourthgen}
In this section we propose a fourth repeater generation as a synthesis of the second and third generations, combining the memory-based architecture of the former with the application of QEC as protection from transmission loss employed in the latter. Even though in this paper we focus on the special case of GKP-based QEC,  where both the stationary and the flying qudits are GKP qudits, the concept of the fourth generation is more general and not restricted to GKP or even bosonic codes.
For a more general discussion on this concept, we refer to App.~\ref{app::sec::generalconcept}.\\

The basic working principle of the fourth generation is very similar to that of the second generation; the total distance is partitioned into segments of length $L_0$ that attempt probabilistic entanglement distribution and store successfully distributed Bell states in quantum memories at the repeater stations. As soon as two neighboring segments both contain entangled states, deterministic entanglement swapping is performed to connect their remote components. 
The feature setting the fourth generation apart from the second generation is the use of GKP encoded states in the entanglement distribution process similar to the third generation.
Specifically, the optical component of the entangled states created locally at the stations will be a GKP-encoded mode instead of multi-rail photons. 
After coupling into the fiber and propagating towards the segment's midpoint, the GKP modes arriving from both directions will be interfered at a balanced beamsplitter and the outputs' quadratures will be measured by homodyne detection. As discussed in Sec.~\ref{sec::GKP}, this implements a Bell state measurement and establishes entanglement between the segment's memory modes.   
Losses incurred during transmission are transformed into random displacements just like for the third generation, resulting in the added variance described in Eq.~(\ref{eq::vardist}), the impact of which can be somewhat mitigated via the HRM. Consequently, the probability for an entanglement distribution attempt to succeed is given by Eq.~(\ref{eq::psucc4}).
In fact, the tradeoff between non-unit success probability and reduced Pauli error probability is a necessity for the concept of the fourth generation to become meaningful, as a repeater with deterministic entanglement distribution would have no need for memories at all, and the scheme would reduce to a third-generation repeater.
Note that we apply the HRM only to entanglement distribution within the segments, but not to entanglement swapping between different segments, which should remain deterministic.

As our fourth-generation repeater is based on the same principle of deterministic entanglement swapping as soon as possible as the second generation, its raw rate can also be calculated in the same manner by inserting the distribution success probability from Eq.~(\ref{eq::psucc4}) into Eq.~(\ref{eq::inverse_raw}) and taking the inverse. 
Determining the QBER, however, is slightly more complex for the fourth generation, due to the presence of an additional source of possible Pauli errors arising from the distribution Bell measurement necessitating a distinction between two Pauli probability vectors $\ppauli{dist}$ and $\ppauli{wait}$ with components given by
\begin{equation}
\left(\ppauli{dist}\right)_r = \frac{\sum_{k \in\mathds{Z}} \int_{(kD + r - h/2)\sqrt{2\pi/D}}^{(kD + r + h/2)\sqrt{2\pi/D}}\, dx \frac{1}{\sqrt{2\pi\chi^2}}\exp\left(\frac{x^2}{2\chi^2}\right)}{\sqrt{p}}
\end{equation}
and 
\begin{equation}
\left(\ppauli{wait}\right)_r = \frac{\sum_{k \in\mathds{Z}} \int_{(kD + r - 1/2)\sqrt{2\pi/D}}^{(kD + r + 1/2)\sqrt{2\pi/D}}\, dx \frac{1}{\sqrt{2\pi\sigma^2}}\exp\left(\frac{x^2}{2\sigma^2}\right)}{\sqrt{p}},
\end{equation}
respectively. Note the integral boundaries' $h$-dependence for $\ppauli{dist}$, as well as the different variances $\chi^2 = 2\delta^2 + \chi^2_\text{add}$ and $\sigma^2 = 2\delta^2 + \mathds{E}(\sigma^2_\text{add})$ arising from transmission loss and memory loss, where $\chi^2_\text{add}$ is defined as before according to Eq.~(\ref{eq::vardist}), and $\mathds{E}(\sigma^2_\text{add})$ is defined by Eq.~(\ref{eq::varwait}), but with $p$ of course now given by Eq.~(\ref{eq::psucc4}).
Further note that we assume identical GKP squeezing as expressed by $\delta^2$ for the transmitted flying and the stored stationary GKP qudits.
The QBER can now be found by combining the Pauli probability vectors through circular convolution with appropriate powers
\begin{equation}
    \boldsymbol{\epsilon} = \left(\ppauli{dist}\right)^{\circledast n}\circledast\left(\ppauli{wait}\right)^{\circledast (n-1)}
\end{equation}
and the SKF by inserting this into Eq.~(\ref{eq::skf}).

It is fairly apparent that the fourth generation can surpass the second only if its distribution success probability $p$ is improved over that of the second; however, this is not a sufficient condition, as even with a better $p$, the decreased SKF due to additional Pauli error sources might negate the raw rate benefits. We expect the fourth generation to be most effective for relatively low coherence times, where shortening average memory wait times has the largest positive effect. 

\section{Comparison of repeater generations}
\label{sec::comparison}
The goal of this section will be to assess each of the repeater generation's performance in terms of SKR; however, for reasons of practical relevance we focus mainly on key rate per unit time instead of per channel use. To account for this, the SKR as calculated following the rate analysis presented in the previous sections must be divided by the duration of an elementary timestep, $\tau_0$.
For memory-based repeaters, $\tau_0$ is governed by the time $L_0/c$ taken up by classical communication across the length of one segment, where $c = 2\times10^{5}$km/s denotes the speed of light in an optical fiber, as well as the state-generation time $\tau_\text{state gen}^\text{mem}$, typically being of the order $10^{-6}$s due to the necessity of light-matter interaction. To simplify matters, we will make the assumption that $\tau_0$ is just given by the maximum of these to time scales, i.e. $\tau_0 = \max(L_0/c, \tau_\text{state gen}^\text{mem})$.
In the case of the memoryless third generation, no classical communication is required, and thus the repetition rate is limited only by state generation and processing. As the generation and entangling of optical GKP states can be achieved without light-matter interactions, e.g. provided that optical cubic phase states are available on demand (see App.~\ref{app::sec::stategen}), we take a value of $\tau_0 = \tau_\text{state gen}^\text{opt} = 10^{-9}$s as a basis for our rate analysis. 
Equipped with an understanding of how to perform rate analyses for our repeater schemes, we now turn towards comparing their performance under various parameter regimes.
A particular point of interest is how the fourth generation fits into the picture, i.e. under which conditions it can offer improved key rates over the established repeater generations.
Unless specifically stated otherwise, we will set $\psucc^\text{(BSM)} = 1/2$ for qubits and $\psucc^\text{(BSM)} = 2/D^2$ in general for the second generation and $\eta_\text{thresh} = 0.1$ for the third and fourth generations. Additionally, the third and fourth generation SKR will be optimized over the HRM acceptance parameter $h$.

\begin{figure}
\begin{center}
\includegraphics[width=\linewidth]{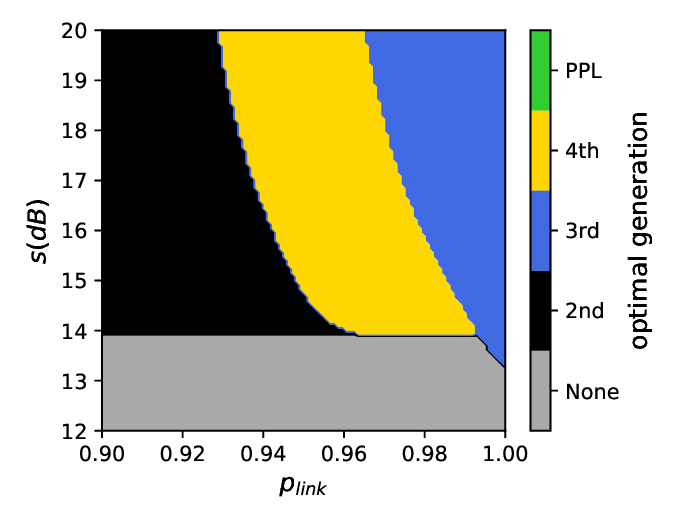}
\end{center}
\caption{(Color online) Optimal repeater generation depending on GKP squeezing and link efficiency (parameters: $D=2$, $n=10000$, $L=1000$km, $\tcoh=1$s). Grey indicates that rates of all generations lie below a threshold of $10^{-5}$Hz, which is the case for squeezing values lower than 14dB. The point-to-point link (PPL) falls short of the threshold for all configurations shown.}
\label{fig::contourplot}
\end{figure}

Figure \ref{fig::contourplot} shows which repeater generation performs best for each point in a plane spanned by GKP squeezing and link efficiency, with the external parameters set at $L=1000$ and $n=10000$km. The second and fourth generations stay below a threshold of $10^{-5}$Hz regardless of the value of $\plink$ as long as the GKP squeezing does not exceed 14dB.
Another noteworthy point is the hierarchy of $\plink$-tolerance, with the third generation exhibiting the lowest, and the second generation the highest tolerance to imperfect couplings, and the fourth generation occupying an intermediate position.
As squeezing is improved, the borders between regions shift towards lower $\plink$, albeit at a decelerating rate, such that even in the infinite squeezing limit, a given value of $\plink$ may not be attainable for the third generation. 

\begin{figure*}
\begin{center}
\includegraphics[width=\linewidth]{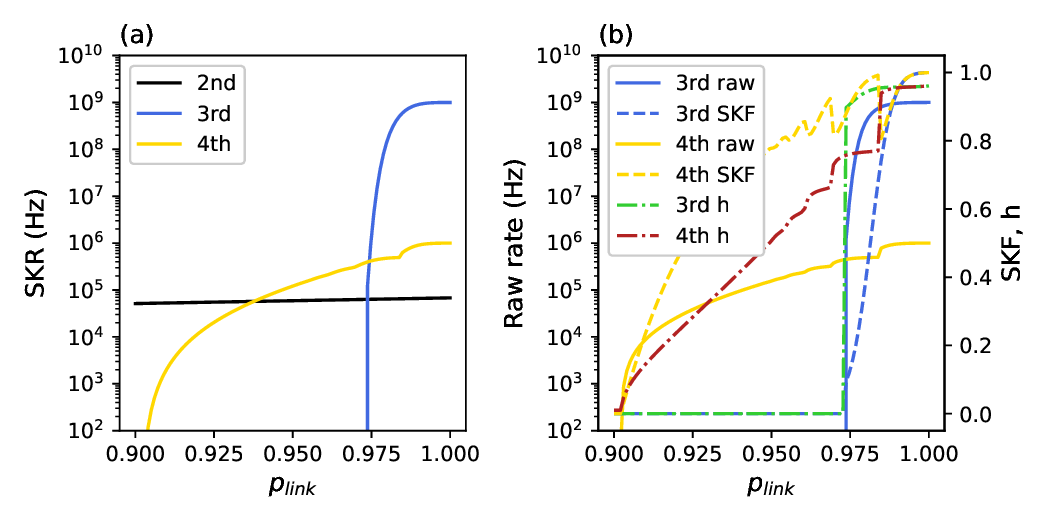}
\end{center}
\caption{(Color online) Cross-section of Fig.~\ref{fig::contourplot} at a squeezing of $s=17$dB (other parameters: $D=2$, $n=10000$, $L=1000$km, $\tcoh=1$s). (a) Secret key rates of all repeater generations (b) Rates split into raw rate and SKF, as well as corresponding optimal HRM parameters, for third and fourth generation. }
\label{fig::contourslice17}
\end{figure*}

Aside from which generation is optimal for a given configuration, it is of course also of interest what rate the optimal generation can achieve, and how big the advantage is over the others. Figure \ref{fig::contourslice17} shows a cross-section of Fig.~\ref{fig::contourplot} at a fixed squeezing of 17dB.  
As noted previously, the second generation is affected the least by decreasing $\plink$, remaining almost constant over the plotted range. 
For sufficiently high $\plink$, the third generation dominates the others by several orders of magnitude due to its higher repetition rate, initially keeping the HRM parameter $h$ close to 1 (at about 0.95) even at the cost of harming the SKF. However, as $\plink$ decreases further to about 0.98, and it becomes necessary to lower $h$ in order to retain a non-zero SKF, the decreasing distribution success probability leads to a rapidly decaying raw rate.
This effect does not occur to the fourth generation, since its raw rate follows the memory-based form of Eq.~(\ref{eq::inverse_raw}) that can tolerate lower distribution success probabilities. Instead, both the SKF and the raw rate reduce more gradually, with $h$ following an almost linear course after some initial irregularities. 

\begin{figure*}
\begin{center}
\includegraphics[width=\linewidth]{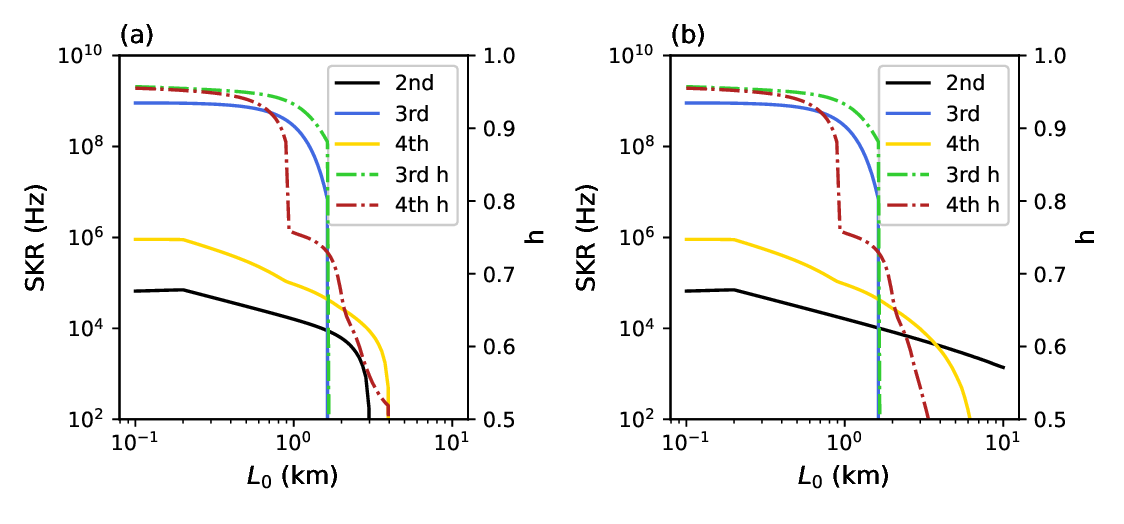}
\end{center}
\caption{(Color online) Rate plotted over segment length for all repeater generations (parameters: $D=2$, $L=1000$km, $s=17$dB, $\plink=0.99$, (a): $\tcoh=1$ms (b): $\tcoh=1$s). For third and fourth generations, the optimal HRM parameters are shown as well. With low memory coherence, the fourth generation performs better than the second for all $L_0$.}
\label{fig::L0plot}
\end{figure*}

In Figs.~\ref{fig::contourplot} and \ref{fig::contourslice17}, the length of the repeater segments was chosen relatively short at $0.1$km; however, for practical applications, it is more desirable to have few intermediate stations and longer segments. To assess how the repeater generations cope with increasingly longer segments, we plot secret key rates over segment length $L_0$ for a fixed total distance of $1000$km in Fig.~\ref{fig::L0plot}, additionally distinguishing between a low-coherence and a high-coherence case for the memory-based generations.
For the second generation, $L_0$ directly affects the distribution success probability, and thus the average memory waiting time, through the exponential relation from Eq.~(\ref{eq::psucc2}). Given a high memory coherence time, longer waiting will result only in a moderately increased state variance, whereas it will quickly surpass the threshold required for a non-zero SKF when the coherence time is low. Therefore, the second generation's behavior at high $L_0$ strongly depends on the coherence time, as witnessed by the fact that the fourth generation achieves higher rates than the second for all $L_0$ at $\tcoh=1$ms, and in particular can yield a non-zero rate even with segment lengths between $2$km and $3$km, where the second generation already vanishes, while at $\tcoh=1$s, the latter can go far beyond $10$km-long segments. 
The third generation's rate decreases the fastest as segment length is increased. This comes down to the same reason as its low $\plink$-tolerance: to counteract the larger variances affecting the Bell state measurements, $h$ must be lowered, leading to a raw rate collapse due to the power-$n$ scaling.  

\begin{figure*}
\begin{center}
\includegraphics[width=\linewidth]{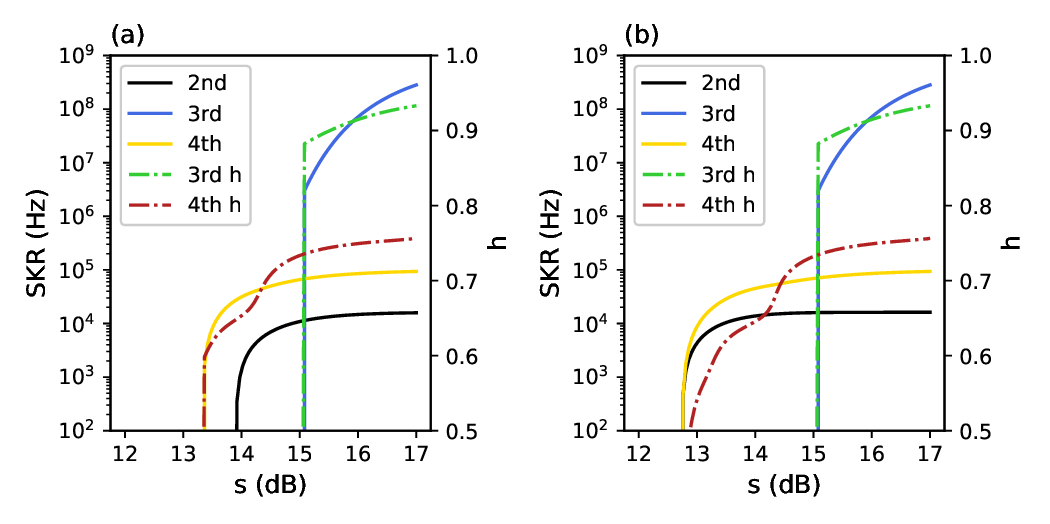}
\end{center}
\caption{(Color online) Rates as a function of GKP squeezing parameter $s$ (other parameters: $D=2$, $L=1000$km, $n=1000$, $\plink=0.99$, (a): $\tcoh=1$ms (b): $\tcoh=1$s). With low memory coherence, the fourth generation requires slightly less squeezing than the second.}
\label{fig::splot}
\end{figure*}

Similar reasoning can also be applied to the third generation's squeezing dependence, which is depicted, together with the other generations', in Fig.~\ref{fig::splot}.  For the chosen parameters, a squeezing of at least $15$dB is required for the third generation, whereas $12.8$dB is sufficient for both the second and fourth generation in the high-coherence case. The thresholds coincide for second and forth, since $\sigma_\text{add}^2$, the contribution to the total variance caused by waiting which differs between generations, is negligible compared to the finite-squeezing variance when the coherence time is high.  
Contrarily, in the low-coherence case, where the average waiting time significantly impacts the rate, the fourth generation's squeezing demand is actually slightly lower than that of the second generation, since the former can select a distribution probability via the HRM that is better than the second generation's, while not being forced to  chose a value so high that it would be damaging to the SKF as the third generation.

\begin{figure*}
\begin{center}
\includegraphics[width=\linewidth]{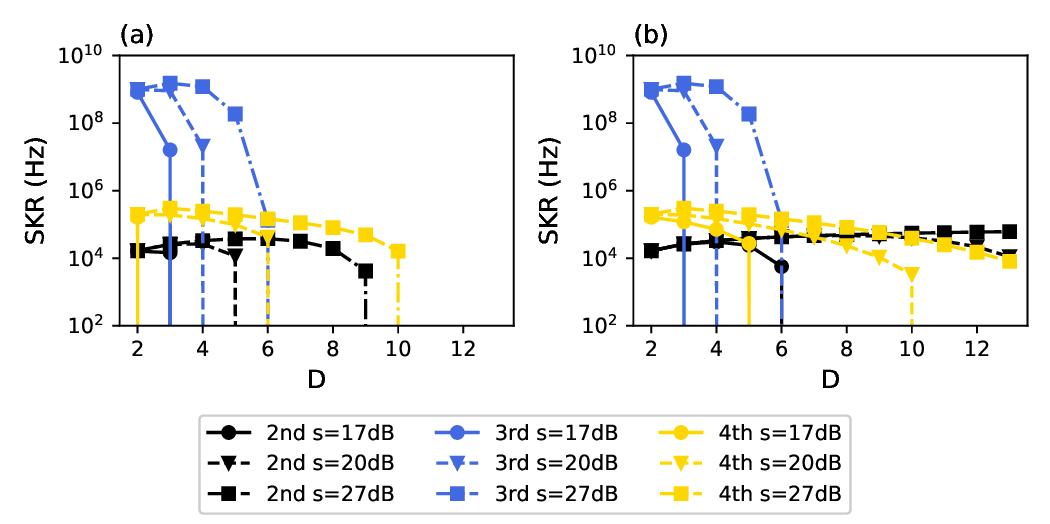}
\end{center}
\caption{(Color online) Rates plotted over the distributed states' logical dimension $D$ for various squeezing levels (parameters: $L=1000$km, $n=1000$, $\plink=1$, (a): $\tcoh=1$ms (b): $\tcoh=1$s). Only for the second generation does a higher dimension lead to a significantly increased rate.}
\label{fig::Dplot}
\end{figure*}

So far we have only considered repeaters using states with logical dimension $D=2$; however, we are also interested in the effect of higher dimensions on the different repeater generations. In Fig.~\ref{fig::Dplot}, we plot rates over $D$ for various levels of squeezing, again distinguishing between coherence times of $1$ms and $1$s.
We observe that only the second generation shows significantly increasing rates for higher $D$, and only at squeezing levels upwards of $20$dB,
whereas third and fourth generation either decrease from the beginning or reach a maximum at $D=3$, depending on the squeezing level.
At low memory coherence, the second generation reaches its maximum at $D=4$ for $s=20$dB and at $D=6$ for $s=27$dB, with the rate increased by about a factor of $2$ compared to the qubit case. Assuming a longer coherence time, the maximum for $s=20$dB lies at $D=8$ and roughly a factor of $3$ above the qubit result, while for $s=27$dB, the rate keeps growing over the entire plotted range, achieving an improvement by a factor of $4$ at $D=13$. However, even with such high-dimensional states, the second generation does not surpass the fourth generation's qubit performance. meaning that the latter is preferable if circumstances allow for it.

\section{Conclusions}
We have introduced a new quantum repeater generation that utilizes QEC to protect both flying states in the entanglement distribution step as well as  stationary states stored in memories at the repeater stations from loss-induced decoherence.   
We have found that, implemented with the GKP encoding, practically relevant parameter regimes exist where the fourth generation offers an advantage over the established second and third generations, in particular the case of short memory coherence times combined with intermediate coupling efficiencies and segment lengths of the order of $1$km.  
If very high coupling efficiency and GKP squeezing are available, and the number of intermediate stations is not too critical, third generation should always be chosen, because it enables secret key rates in the range of $10^7$Hz to $10^9$Hz that are unattainable for the memory-based generations. 
If, however, the necessary high $\plink$ values cannot be reached, but memories with relatively long coherence times are available, one is presented with the choice between operating at lower rates but with fewer repeater stations as enabled by the second generation, or maintaining higher rates with more stations using the fourth generation; depending on whether cost or performance is the critical factor. 
The modification of the repeater protocols to use higher-dimensional quantum states appears to be of rather limited usefulness, however. While possible for all generations if the GKP squeezing is sufficient, significant gains are found only for the second generation. Thus we conclude that the use of qudits instead of qubits is only beneficial when the parameters in a given application context do not allow for third or fourth generation repeaters.

\bibliography{literatur}

\begin{thebibliography}{44}%
\makeatletter
\providecommand \@ifxundefined [1]{%
 \@ifx{#1\undefined}
}%
\providecommand \@ifnum [1]{%
 \ifnum #1\expandafter \@firstoftwo
 \else \expandafter \@secondoftwo
 \fi
}%
\providecommand \@ifx [1]{%
 \ifx #1\expandafter \@firstoftwo
 \else \expandafter \@secondoftwo
 \fi
}%
\providecommand \natexlab [1]{#1}%
\providecommand \enquote  [1]{``#1''}%
\providecommand \bibnamefont  [1]{#1}%
\providecommand \bibfnamefont [1]{#1}%
\providecommand \citenamefont [1]{#1}%
\providecommand \href@noop [0]{\@secondoftwo}%
\providecommand \href [0]{\begingroup \@sanitize@url \@href}%
\providecommand \@href[1]{\@@startlink{#1}\@@href}%
\providecommand \@@href[1]{\endgroup#1\@@endlink}%
\providecommand \@sanitize@url [0]{\catcode `\\12\catcode `\$12\catcode `\&12\catcode `\#12\catcode `\^12\catcode `\_12\catcode `\%12\relax}%
\providecommand \@@startlink[1]{}%
\providecommand \@@endlink[0]{}%
\providecommand \url  [0]{\begingroup\@sanitize@url \@url }%
\providecommand \@url [1]{\endgroup\@href {#1}{\urlprefix }}%
\providecommand \urlprefix  [0]{URL }%
\providecommand \Eprint [0]{\href }%
\providecommand \doibase [0]{http://dx.doi.org/}%
\providecommand \selectlanguage [0]{\@gobble}%
\providecommand \bibinfo  [0]{\@secondoftwo}%
\providecommand \bibfield  [0]{\@secondoftwo}%
\providecommand \translation [1]{[#1]}%
\providecommand \BibitemOpen [0]{}%
\providecommand \bibitemStop [0]{}%
\providecommand \bibitemNoStop [0]{.\EOS\space}%
\providecommand \EOS [0]{\spacefactor3000\relax}%
\providecommand \BibitemShut  [1]{\csname bibitem#1\endcsname}%
\let\auto@bib@innerbib\@empty
\bibitem [{\citenamefont {Bennett}\ and\ \citenamefont {Brassard}(1984)}]{BB84}%
  \BibitemOpen
  \bibfield  {author} {\bibinfo {author} {\bibfnamefont {C.}~\bibnamefont {Bennett}}\ and\ \bibinfo {author} {\bibfnamefont {G.}~\bibnamefont {Brassard}}\ }(\bibinfo {year} {1984})\ pp.\ \bibinfo {pages} {175--179}\BibitemShut {NoStop}%
\bibitem [{\citenamefont {Cuomo}\ \emph {et~al.}(2020)\citenamefont {Cuomo}, \citenamefont {Caleffi},\ and\ \citenamefont {Cacciapuoti}}]{eco}%
  \BibitemOpen
  \bibfield  {author} {\bibinfo {author} {\bibfnamefont {D.}~\bibnamefont {Cuomo}}, \bibinfo {author} {\bibfnamefont {M.}~\bibnamefont {Caleffi}}, \ and\ \bibinfo {author} {\bibfnamefont {A.~S.}\ \bibnamefont {Cacciapuoti}},\ }\href {\doibase https://doi.org/10.1049/iet-qtc.2020.0002} {\bibfield  {journal} {\bibinfo  {journal} {IET Quantum Communication}\ }\textbf {\bibinfo {volume} {1}},\ \bibinfo {pages} {3} (\bibinfo {year} {2020})}\BibitemShut {NoStop}%
\bibitem [{\citenamefont {Takeoka}\ \emph {et~al.}(2014)\citenamefont {Takeoka}, \citenamefont {Guha},\ and\ \citenamefont {Wilde}}]{Takeoka2014}%
  \BibitemOpen
  \bibfield  {author} {\bibinfo {author} {\bibfnamefont {M.}~\bibnamefont {Takeoka}}, \bibinfo {author} {\bibfnamefont {S.}~\bibnamefont {Guha}}, \ and\ \bibinfo {author} {\bibfnamefont {M.~M.}\ \bibnamefont {Wilde}},\ }\href {\doibase 10.1038/ncomms6235} {\bibfield  {journal} {\bibinfo  {journal} {Nature Communications}\ }\textbf {\bibinfo {volume} {5}},\ \bibinfo {pages} {5235} (\bibinfo {year} {2014})}\BibitemShut {NoStop}%
\bibitem [{\citenamefont {Pirandola}\ \emph {et~al.}(2017)\citenamefont {Pirandola}, \citenamefont {Laurenza}, \citenamefont {Ottaviani},\ and\ \citenamefont {Banchi}}]{plob}%
  \BibitemOpen
  \bibfield  {author} {\bibinfo {author} {\bibfnamefont {S.}~\bibnamefont {Pirandola}}, \bibinfo {author} {\bibfnamefont {R.}~\bibnamefont {Laurenza}}, \bibinfo {author} {\bibfnamefont {C.}~\bibnamefont {Ottaviani}}, \ and\ \bibinfo {author} {\bibfnamefont {L.}~\bibnamefont {Banchi}},\ }\href {\doibase 10.1038/ncomms15043} {\bibfield  {journal} {\bibinfo  {journal} {Nature Communications}\ }\textbf {\bibinfo {volume} {8}},\ \bibinfo {pages} {15043} (\bibinfo {year} {2017})}\BibitemShut {NoStop}%
\bibitem [{\citenamefont {Pirandola}(2019)}]{repeaterassisted}%
  \BibitemOpen
  \bibfield  {author} {\bibinfo {author} {\bibfnamefont {S.}~\bibnamefont {Pirandola}},\ }\href {\doibase 10.1038/s42005-019-0147-3} {\bibfield  {journal} {\bibinfo  {journal} {Communications Physics}\ }\textbf {\bibinfo {volume} {2}},\ \bibinfo {pages} {51} (\bibinfo {year} {2019})}\BibitemShut {NoStop}%
\bibitem [{\citenamefont {Briegel}\ \emph {et~al.}(1998)\citenamefont {Briegel}, \citenamefont {D\"ur}, \citenamefont {Cirac},\ and\ \citenamefont {Zoller}}]{firstrepeater}%
  \BibitemOpen
  \bibfield  {author} {\bibinfo {author} {\bibfnamefont {H.-J.}\ \bibnamefont {Briegel}}, \bibinfo {author} {\bibfnamefont {W.}~\bibnamefont {D\"ur}}, \bibinfo {author} {\bibfnamefont {J.~I.}\ \bibnamefont {Cirac}}, \ and\ \bibinfo {author} {\bibfnamefont {P.}~\bibnamefont {Zoller}},\ }\href {\doibase 10.1103/PhysRevLett.81.5932} {\bibfield  {journal} {\bibinfo  {journal} {Phys. Rev. Lett.}\ }\textbf {\bibinfo {volume} {81}},\ \bibinfo {pages} {5932} (\bibinfo {year} {1998})}\BibitemShut {NoStop}%
\bibitem [{\citenamefont {Muralidharan}\ \emph {et~al.}(2016)\citenamefont {Muralidharan}, \citenamefont {Li}, \citenamefont {Kim}, \citenamefont {Lütkenhaus}, \citenamefont {Lukin},\ and\ \citenamefont {Jiang}}]{generations}%
  \BibitemOpen
  \bibfield  {author} {\bibinfo {author} {\bibfnamefont {S.}~\bibnamefont {Muralidharan}}, \bibinfo {author} {\bibfnamefont {L.}~\bibnamefont {Li}}, \bibinfo {author} {\bibfnamefont {J.}~\bibnamefont {Kim}}, \bibinfo {author} {\bibfnamefont {N.}~\bibnamefont {Lütkenhaus}}, \bibinfo {author} {\bibfnamefont {M.~D.}\ \bibnamefont {Lukin}}, \ and\ \bibinfo {author} {\bibfnamefont {L.}~\bibnamefont {Jiang}},\ }\href {\doibase 10.1038/srep20463} {\bibfield  {journal} {\bibinfo  {journal} {Scientific Reports}\ }\textbf {\bibinfo {volume} {6}},\ \bibinfo {pages} {20463} (\bibinfo {year} {2016})}\BibitemShut {NoStop}%
\bibitem [{\citenamefont {Jiang}\ \emph {et~al.}(2009)\citenamefont {Jiang}, \citenamefont {Taylor}, \citenamefont {Nemoto}, \citenamefont {Munro}, \citenamefont {Van~Meter},\ and\ \citenamefont {Lukin}}]{secondrepeater}%
  \BibitemOpen
  \bibfield  {author} {\bibinfo {author} {\bibfnamefont {L.}~\bibnamefont {Jiang}}, \bibinfo {author} {\bibfnamefont {J.~M.}\ \bibnamefont {Taylor}}, \bibinfo {author} {\bibfnamefont {K.}~\bibnamefont {Nemoto}}, \bibinfo {author} {\bibfnamefont {W.~J.}\ \bibnamefont {Munro}}, \bibinfo {author} {\bibfnamefont {R.}~\bibnamefont {Van~Meter}}, \ and\ \bibinfo {author} {\bibfnamefont {M.~D.}\ \bibnamefont {Lukin}},\ }\href {\doibase 10.1103/PhysRevA.79.032325} {\bibfield  {journal} {\bibinfo  {journal} {Phys. Rev. A}\ }\textbf {\bibinfo {volume} {79}},\ \bibinfo {pages} {032325} (\bibinfo {year} {2009})}\BibitemShut {NoStop}%
\bibitem [{\citenamefont {Fowler}\ \emph {et~al.}(2010)\citenamefont {Fowler}, \citenamefont {Wang}, \citenamefont {Hill}, \citenamefont {Ladd}, \citenamefont {Van~Meter},\ and\ \citenamefont {Hollenberg}}]{thirdgen_surfacecode}%
  \BibitemOpen
  \bibfield  {author} {\bibinfo {author} {\bibfnamefont {A.~G.}\ \bibnamefont {Fowler}}, \bibinfo {author} {\bibfnamefont {D.~S.}\ \bibnamefont {Wang}}, \bibinfo {author} {\bibfnamefont {C.~D.}\ \bibnamefont {Hill}}, \bibinfo {author} {\bibfnamefont {T.~D.}\ \bibnamefont {Ladd}}, \bibinfo {author} {\bibfnamefont {R.}~\bibnamefont {Van~Meter}}, \ and\ \bibinfo {author} {\bibfnamefont {L.~C.~L.}\ \bibnamefont {Hollenberg}},\ }\href {\doibase 10.1103/PhysRevLett.104.180503} {\bibfield  {journal} {\bibinfo  {journal} {Phys. Rev. Lett.}\ }\textbf {\bibinfo {volume} {104}},\ \bibinfo {pages} {180503} (\bibinfo {year} {2010})}\BibitemShut {NoStop}%
\bibitem [{\citenamefont {Munro}\ \emph {et~al.}(2012)\citenamefont {Munro}, \citenamefont {Stephens}, \citenamefont {Devitt}, \citenamefont {Harrison},\ and\ \citenamefont {Nemoto}}]{Munro2012}%
  \BibitemOpen
  \bibfield  {author} {\bibinfo {author} {\bibfnamefont {W.~J.}\ \bibnamefont {Munro}}, \bibinfo {author} {\bibfnamefont {A.~M.}\ \bibnamefont {Stephens}}, \bibinfo {author} {\bibfnamefont {S.~J.}\ \bibnamefont {Devitt}}, \bibinfo {author} {\bibfnamefont {K.~A.}\ \bibnamefont {Harrison}}, \ and\ \bibinfo {author} {\bibfnamefont {K.}~\bibnamefont {Nemoto}},\ }\href {\doibase 10.1038/nphoton.2012.243} {\bibfield  {journal} {\bibinfo  {journal} {Nature Photonics}\ }\textbf {\bibinfo {volume} {6}},\ \bibinfo {pages} {777} (\bibinfo {year} {2012})}\BibitemShut {NoStop}%
\bibitem [{\citenamefont {Azuma}\ \emph {et~al.}(2015)\citenamefont {Azuma}, \citenamefont {Tamaki},\ and\ \citenamefont {Lo}}]{Azuma2015}%
  \BibitemOpen
  \bibfield  {author} {\bibinfo {author} {\bibfnamefont {K.}~\bibnamefont {Azuma}}, \bibinfo {author} {\bibfnamefont {K.}~\bibnamefont {Tamaki}}, \ and\ \bibinfo {author} {\bibfnamefont {H.-K.}\ \bibnamefont {Lo}},\ }\href {\doibase 10.1038/ncomms7787} {\bibfield  {journal} {\bibinfo  {journal} {Nature Communications}\ }\textbf {\bibinfo {volume} {6}},\ \bibinfo {pages} {6787} (\bibinfo {year} {2015})}\BibitemShut {NoStop}%
\bibitem [{\citenamefont {Muralidharan}\ \emph {et~al.}(2014)\citenamefont {Muralidharan}, \citenamefont {Kim}, \citenamefont {L\"utkenhaus}, \citenamefont {Lukin},\ and\ \citenamefont {Jiang}}]{Muralidharan2014}%
  \BibitemOpen
  \bibfield  {author} {\bibinfo {author} {\bibfnamefont {S.}~\bibnamefont {Muralidharan}}, \bibinfo {author} {\bibfnamefont {J.}~\bibnamefont {Kim}}, \bibinfo {author} {\bibfnamefont {N.}~\bibnamefont {L\"utkenhaus}}, \bibinfo {author} {\bibfnamefont {M.~D.}\ \bibnamefont {Lukin}}, \ and\ \bibinfo {author} {\bibfnamefont {L.}~\bibnamefont {Jiang}},\ }\href {\doibase 10.1103/PhysRevLett.112.250501} {\bibfield  {journal} {\bibinfo  {journal} {Phys. Rev. Lett.}\ }\textbf {\bibinfo {volume} {112}},\ \bibinfo {pages} {250501} (\bibinfo {year} {2014})}\BibitemShut {NoStop}%
\bibitem [{\citenamefont {Ewert}\ \emph {et~al.}(2016)\citenamefont {Ewert}, \citenamefont {Bergmann},\ and\ \citenamefont {van Loock}}]{Ewert2016}%
  \BibitemOpen
  \bibfield  {author} {\bibinfo {author} {\bibfnamefont {F.}~\bibnamefont {Ewert}}, \bibinfo {author} {\bibfnamefont {M.}~\bibnamefont {Bergmann}}, \ and\ \bibinfo {author} {\bibfnamefont {P.}~\bibnamefont {van Loock}},\ }\href {\doibase 10.1103/PhysRevLett.117.210501} {\bibfield  {journal} {\bibinfo  {journal} {Phys. Rev. Lett.}\ }\textbf {\bibinfo {volume} {117}},\ \bibinfo {pages} {210501} (\bibinfo {year} {2016})}\BibitemShut {NoStop}%
\bibitem [{\citenamefont {Ewert}\ and\ \citenamefont {van Loock}(2017)}]{Ewert2017}%
  \BibitemOpen
  \bibfield  {author} {\bibinfo {author} {\bibfnamefont {F.}~\bibnamefont {Ewert}}\ and\ \bibinfo {author} {\bibfnamefont {P.}~\bibnamefont {van Loock}},\ }\href {\doibase 10.1103/PhysRevA.95.012327} {\bibfield  {journal} {\bibinfo  {journal} {Phys. Rev. A}\ }\textbf {\bibinfo {volume} {95}},\ \bibinfo {pages} {012327} (\bibinfo {year} {2017})}\BibitemShut {NoStop}%
\bibitem [{\citenamefont {Lee}\ \emph {et~al.}(2019)\citenamefont {Lee}, \citenamefont {Ralph},\ and\ \citenamefont {Jeong}}]{Lee2019}%
  \BibitemOpen
  \bibfield  {author} {\bibinfo {author} {\bibfnamefont {S.-W.}\ \bibnamefont {Lee}}, \bibinfo {author} {\bibfnamefont {T.~C.}\ \bibnamefont {Ralph}}, \ and\ \bibinfo {author} {\bibfnamefont {H.}~\bibnamefont {Jeong}},\ }\href {\doibase 10.1103/PhysRevA.100.052303} {\bibfield  {journal} {\bibinfo  {journal} {Phys. Rev. A}\ }\textbf {\bibinfo {volume} {100}},\ \bibinfo {pages} {052303} (\bibinfo {year} {2019})}\BibitemShut {NoStop}%
\bibitem [{\citenamefont {Gottesman}\ \emph {et~al.}(2001)\citenamefont {Gottesman}, \citenamefont {Kitaev},\ and\ \citenamefont {Preskill}}]{gkp}%
  \BibitemOpen
  \bibfield  {author} {\bibinfo {author} {\bibfnamefont {D.}~\bibnamefont {Gottesman}}, \bibinfo {author} {\bibfnamefont {A.}~\bibnamefont {Kitaev}}, \ and\ \bibinfo {author} {\bibfnamefont {J.}~\bibnamefont {Preskill}},\ }\href {\doibase 10.1103/PhysRevA.64.012310} {\bibfield  {journal} {\bibinfo  {journal} {Phys. Rev. A}\ }\textbf {\bibinfo {volume} {64}},\ \bibinfo {pages} {012310} (\bibinfo {year} {2001})}\BibitemShut {NoStop}%
\bibitem [{\citenamefont {H\"aussler}\ and\ \citenamefont {van Loock}(2025)}]{häussler_vanloock}%
  \BibitemOpen
  \bibfield  {author} {\bibinfo {author} {\bibfnamefont {S.}~\bibnamefont {H\"aussler}}\ and\ \bibinfo {author} {\bibfnamefont {P.}~\bibnamefont {van Loock}},\ }\href {\doibase 10.1103/PhysRevA.111.062611} {\bibfield  {journal} {\bibinfo  {journal} {Phys. Rev. A}\ }\textbf {\bibinfo {volume} {111}},\ \bibinfo {pages} {062611} (\bibinfo {year} {2025})}\BibitemShut {NoStop}%
\bibitem [{\citenamefont {Schmidt}\ \emph {et~al.}(2024)\citenamefont {Schmidt}, \citenamefont {Miller},\ and\ \citenamefont {van Loock}}]{thirdgen_gkp}%
  \BibitemOpen
  \bibfield  {author} {\bibinfo {author} {\bibfnamefont {F.}~\bibnamefont {Schmidt}}, \bibinfo {author} {\bibfnamefont {D.}~\bibnamefont {Miller}}, \ and\ \bibinfo {author} {\bibfnamefont {P.}~\bibnamefont {van Loock}},\ }\href {\doibase 10.1103/PhysRevA.109.042427} {\bibfield  {journal} {\bibinfo  {journal} {Phys. Rev. A}\ }\textbf {\bibinfo {volume} {109}},\ \bibinfo {pages} {042427} (\bibinfo {year} {2024})}\BibitemShut {NoStop}%
\bibitem [{\citenamefont {Fukui}\ \emph {et~al.}(2021)\citenamefont {Fukui}, \citenamefont {Alexander},\ and\ \citenamefont {van Loock}}]{alloptical}%
  \BibitemOpen
  \bibfield  {author} {\bibinfo {author} {\bibfnamefont {K.}~\bibnamefont {Fukui}}, \bibinfo {author} {\bibfnamefont {R.~N.}\ \bibnamefont {Alexander}}, \ and\ \bibinfo {author} {\bibfnamefont {P.}~\bibnamefont {van Loock}},\ }\href {\doibase 10.1103/PhysRevResearch.3.033118} {\bibfield  {journal} {\bibinfo  {journal} {Phys. Rev. Res.}\ }\textbf {\bibinfo {volume} {3}},\ \bibinfo {pages} {033118} (\bibinfo {year} {2021})}\BibitemShut {NoStop}%
\bibitem [{\citenamefont {Rozp{\k{e}}dek}\ \emph {et~al.}(2021)\citenamefont {Rozp{\k{e}}dek}, \citenamefont {Noh}, \citenamefont {Xu}, \citenamefont {Guha},\ and\ \citenamefont {Jiang}}]{Rozpedek2021}%
  \BibitemOpen
  \bibfield  {author} {\bibinfo {author} {\bibfnamefont {F.}~\bibnamefont {Rozp{\k{e}}dek}}, \bibinfo {author} {\bibfnamefont {K.}~\bibnamefont {Noh}}, \bibinfo {author} {\bibfnamefont {Q.}~\bibnamefont {Xu}}, \bibinfo {author} {\bibfnamefont {S.}~\bibnamefont {Guha}}, \ and\ \bibinfo {author} {\bibfnamefont {L.}~\bibnamefont {Jiang}},\ }\href {\doibase 10.1038/s41534-021-00438-7} {\bibfield  {journal} {\bibinfo  {journal} {npj Quantum Information}\ }\textbf {\bibinfo {volume} {7}},\ \bibinfo {pages} {102} (\bibinfo {year} {2021})}\BibitemShut {NoStop}%
\bibitem [{\citenamefont {Rozp\ifmmode~\mbox{\k{e}}\else \k{e}\fi{}dek}\ \emph {et~al.}(2023)\citenamefont {Rozp\ifmmode~\mbox{\k{e}}\else \k{e}\fi{}dek}, \citenamefont {Seshadreesan}, \citenamefont {Polakos}, \citenamefont {Jiang},\ and\ \citenamefont {Guha}}]{Rozpedek2023}%
  \BibitemOpen
  \bibfield  {author} {\bibinfo {author} {\bibfnamefont {F.}~\bibnamefont {Rozp\ifmmode~\mbox{\k{e}}\else \k{e}\fi{}dek}}, \bibinfo {author} {\bibfnamefont {K.~P.}\ \bibnamefont {Seshadreesan}}, \bibinfo {author} {\bibfnamefont {P.}~\bibnamefont {Polakos}}, \bibinfo {author} {\bibfnamefont {L.}~\bibnamefont {Jiang}}, \ and\ \bibinfo {author} {\bibfnamefont {S.}~\bibnamefont {Guha}},\ }\href {\doibase 10.1103/PhysRevResearch.5.043056} {\bibfield  {journal} {\bibinfo  {journal} {Phys. Rev. Res.}\ }\textbf {\bibinfo {volume} {5}},\ \bibinfo {pages} {043056} (\bibinfo {year} {2023})}\BibitemShut {NoStop}%
\bibitem [{\citenamefont {Hammerer}\ \emph {et~al.}(2010)\citenamefont {Hammerer}, \citenamefont {S\o{}rensen},\ and\ \citenamefont {Polzik}}]{hammerer}%
  \BibitemOpen
  \bibfield  {author} {\bibinfo {author} {\bibfnamefont {K.}~\bibnamefont {Hammerer}}, \bibinfo {author} {\bibfnamefont {A.~S.}\ \bibnamefont {S\o{}rensen}}, \ and\ \bibinfo {author} {\bibfnamefont {E.~S.}\ \bibnamefont {Polzik}},\ }\href {\doibase 10.1103/RevModPhys.82.1041} {\bibfield  {journal} {\bibinfo  {journal} {Rev. Mod. Phys.}\ }\textbf {\bibinfo {volume} {82}},\ \bibinfo {pages} {1041} (\bibinfo {year} {2010})}\BibitemShut {NoStop}%
\bibitem [{\citenamefont {Omanakuttan}\ and\ \citenamefont {Volkoff}(2023)}]{ensgkp}%
  \BibitemOpen
  \bibfield  {author} {\bibinfo {author} {\bibfnamefont {S.}~\bibnamefont {Omanakuttan}}\ and\ \bibinfo {author} {\bibfnamefont {T.~J.}\ \bibnamefont {Volkoff}},\ }\href {\doibase 10.1103/PhysRevA.108.022428} {\bibfield  {journal} {\bibinfo  {journal} {Phys. Rev. A}\ }\textbf {\bibinfo {volume} {108}},\ \bibinfo {pages} {022428} (\bibinfo {year} {2023})}\BibitemShut {NoStop}%
\bibitem [{\citenamefont {Holstein}\ and\ \citenamefont {Primakoff}(1940)}]{hp}%
  \BibitemOpen
  \bibfield  {author} {\bibinfo {author} {\bibfnamefont {T.}~\bibnamefont {Holstein}}\ and\ \bibinfo {author} {\bibfnamefont {H.}~\bibnamefont {Primakoff}},\ }\href {\doibase 10.1103/PhysRev.58.1098} {\bibfield  {journal} {\bibinfo  {journal} {Phys. Rev.}\ }\textbf {\bibinfo {volume} {58}},\ \bibinfo {pages} {1098} (\bibinfo {year} {1940})}\BibitemShut {NoStop}%
\bibitem [{\citenamefont {Calsamiglia}\ and\ \citenamefont {L{\"u}tkenhaus}(2001)}]{Calsamiglia2001}%
  \BibitemOpen
  \bibfield  {author} {\bibinfo {author} {\bibfnamefont {J.}~\bibnamefont {Calsamiglia}}\ and\ \bibinfo {author} {\bibfnamefont {N.}~\bibnamefont {L{\"u}tkenhaus}},\ }\href {\doibase 10.1007/s003400000484} {\bibfield  {journal} {\bibinfo  {journal} {Applied Physics B}\ }\textbf {\bibinfo {volume} {72}},\ \bibinfo {pages} {67} (\bibinfo {year} {2001})}\BibitemShut {NoStop}%
\bibitem [{\citenamefont {Knill}\ \emph {et~al.}(2001)\citenamefont {Knill}, \citenamefont {Laflamme},\ and\ \citenamefont {Milburn}}]{Knill2001}%
  \BibitemOpen
  \bibfield  {author} {\bibinfo {author} {\bibfnamefont {E.}~\bibnamefont {Knill}}, \bibinfo {author} {\bibfnamefont {R.}~\bibnamefont {Laflamme}}, \ and\ \bibinfo {author} {\bibfnamefont {G.~J.}\ \bibnamefont {Milburn}},\ }\href {\doibase 10.1038/35051009} {\bibfield  {journal} {\bibinfo  {journal} {Nature}\ }\textbf {\bibinfo {volume} {409}},\ \bibinfo {pages} {46} (\bibinfo {year} {2001})}\BibitemShut {NoStop}%
\bibitem [{\citenamefont {Grice}(2011)}]{Grice2011}%
  \BibitemOpen
  \bibfield  {author} {\bibinfo {author} {\bibfnamefont {W.~P.}\ \bibnamefont {Grice}},\ }\href {\doibase 10.1103/PhysRevA.84.042331} {\bibfield  {journal} {\bibinfo  {journal} {Phys. Rev. A}\ }\textbf {\bibinfo {volume} {84}},\ \bibinfo {pages} {042331} (\bibinfo {year} {2011})}\BibitemShut {NoStop}%
\bibitem [{\citenamefont {Ewert}\ and\ \citenamefont {van Loock}(2014)}]{Ewert_bsm}%
  \BibitemOpen
  \bibfield  {author} {\bibinfo {author} {\bibfnamefont {F.}~\bibnamefont {Ewert}}\ and\ \bibinfo {author} {\bibfnamefont {P.}~\bibnamefont {van Loock}},\ }\href {\doibase 10.1103/PhysRevLett.113.140403} {\bibfield  {journal} {\bibinfo  {journal} {Phys. Rev. Lett.}\ }\textbf {\bibinfo {volume} {113}},\ \bibinfo {pages} {140403} (\bibinfo {year} {2014})}\BibitemShut {NoStop}%
\bibitem [{\citenamefont {Bayerbach}\ \emph {et~al.}(2023)\citenamefont {Bayerbach}, \citenamefont {D’Aurelio}, \citenamefont {van Loock},\ and\ \citenamefont {Barz}}]{Bayerbach2023}%
  \BibitemOpen
  \bibfield  {author} {\bibinfo {author} {\bibfnamefont {M.~J.}\ \bibnamefont {Bayerbach}}, \bibinfo {author} {\bibfnamefont {S.~E.}\ \bibnamefont {D’Aurelio}}, \bibinfo {author} {\bibfnamefont {P.}~\bibnamefont {van Loock}}, \ and\ \bibinfo {author} {\bibfnamefont {S.}~\bibnamefont {Barz}},\ }\href {\doibase 10.1126/sciadv.adf4080} {\bibfield  {journal} {\bibinfo  {journal} {Science Advances}\ }\textbf {\bibinfo {volume} {9}},\ \bibinfo {pages} {eadf4080} (\bibinfo {year} {2023})},\ \Eprint {http://arxiv.org/abs/https://www.science.org/doi/pdf/10.1126/sciadv.adf4080} {https://www.science.org/doi/pdf/10.1126/sciadv.adf4080} \BibitemShut {NoStop}%
\bibitem [{\citenamefont {Guo}\ \emph {et~al.}(2024)\citenamefont {Guo}, \citenamefont {Zou}, \citenamefont {Ding}, \citenamefont {Zhang}, \citenamefont {Xu}, \citenamefont {Liu}, \citenamefont {Zhao}, \citenamefont {Ge}, \citenamefont {Peng}, \citenamefont {Xu}, \citenamefont {Lou}, \citenamefont {Ning}, \citenamefont {Wang}, \citenamefont {Wang}, \citenamefont {Huo}, \citenamefont {He}, \citenamefont {Lu},\ and\ \citenamefont {Pan}}]{boostedfusiongates}%
  \BibitemOpen
  \bibfield  {author} {\bibinfo {author} {\bibfnamefont {Y.-P.}\ \bibnamefont {Guo}}, \bibinfo {author} {\bibfnamefont {G.-Y.}\ \bibnamefont {Zou}}, \bibinfo {author} {\bibfnamefont {X.}~\bibnamefont {Ding}}, \bibinfo {author} {\bibfnamefont {Q.-H.}\ \bibnamefont {Zhang}}, \bibinfo {author} {\bibfnamefont {M.-C.}\ \bibnamefont {Xu}}, \bibinfo {author} {\bibfnamefont {R.-Z.}\ \bibnamefont {Liu}}, \bibinfo {author} {\bibfnamefont {J.-Y.}\ \bibnamefont {Zhao}}, \bibinfo {author} {\bibfnamefont {Z.-X.}\ \bibnamefont {Ge}}, \bibinfo {author} {\bibfnamefont {L.-C.}\ \bibnamefont {Peng}}, \bibinfo {author} {\bibfnamefont {K.-M.}\ \bibnamefont {Xu}}, \bibinfo {author} {\bibfnamefont {Y.-Y.}\ \bibnamefont {Lou}}, \bibinfo {author} {\bibfnamefont {Z.}~\bibnamefont {Ning}}, \bibinfo {author} {\bibfnamefont {L.-J.}\ \bibnamefont {Wang}}, \bibinfo {author} {\bibfnamefont {H.}~\bibnamefont {Wang}}, \bibinfo {author} {\bibfnamefont {Y.-H.}\ \bibnamefont {Huo}}, \bibinfo {author} {\bibfnamefont {Y.-M.}\ \bibnamefont {He}},
  \bibinfo {author} {\bibfnamefont {C.-Y.}\ \bibnamefont {Lu}}, \ and\ \bibinfo {author} {\bibfnamefont {J.-W.}\ \bibnamefont {Pan}},\ }\href {https://arxiv.org/abs/2412.18882} {\enquote {\bibinfo {title} {Boosted fusion gates above the percolation threshold for scalable graph-state generation},}\ } (\bibinfo {year} {2024}),\ \Eprint {http://arxiv.org/abs/2412.18882} {arXiv:2412.18882 [quant-ph]} \BibitemShut {NoStop}%
\bibitem [{\citenamefont {Calsamiglia}(2002)}]{Calsamiglia2002}%
  \BibitemOpen
  \bibfield  {author} {\bibinfo {author} {\bibfnamefont {J.}~\bibnamefont {Calsamiglia}},\ }\href {\doibase 10.1103/PhysRevA.65.030301} {\bibfield  {journal} {\bibinfo  {journal} {Phys. Rev. A}\ }\textbf {\bibinfo {volume} {65}},\ \bibinfo {pages} {030301} (\bibinfo {year} {2002})}\BibitemShut {NoStop}%
\bibitem [{\citenamefont {Carollo}\ and\ \citenamefont {Palma}(2001)}]{Carollo}%
  \BibitemOpen
  \bibfield  {author} {\bibinfo {author} {\bibfnamefont {A.}~\bibnamefont {Carollo}}\ and\ \bibinfo {author} {\bibfnamefont {G.~M.}\ \bibnamefont {Palma}},\ }in\ \href {https://opg.optica.org/abstract.cfm?URI=CLEO_Europe-2001-QC281} {\emph {\bibinfo {booktitle} {Laser 2001 - World of Photonics 15th International Conference on Lasers and Electrooptics in Europe}}}\ (\bibinfo  {publisher} {Optica Publishing Group},\ \bibinfo {year} {2001})\ p.\ \bibinfo {pages} {QC281}\BibitemShut {NoStop}%
\bibitem [{\citenamefont {Dušek}(2001)}]{Dusek}%
  \BibitemOpen
  \bibfield  {author} {\bibinfo {author} {\bibfnamefont {M.}~\bibnamefont {Dušek}},\ }\href {\doibase https://doi.org/10.1016/S0030-4018(01)01565-6} {\bibfield  {journal} {\bibinfo  {journal} {Optics Communications}\ }\textbf {\bibinfo {volume} {199}},\ \bibinfo {pages} {161} (\bibinfo {year} {2001})}\BibitemShut {NoStop}%
\bibitem [{\citenamefont {Bharos}\ \emph {et~al.}(2024)\citenamefont {Bharos}, \citenamefont {Markovich},\ and\ \citenamefont {Borregaard}}]{Bharos}%
  \BibitemOpen
  \bibfield  {author} {\bibinfo {author} {\bibfnamefont {N.}~\bibnamefont {Bharos}}, \bibinfo {author} {\bibfnamefont {L.}~\bibnamefont {Markovich}}, \ and\ \bibinfo {author} {\bibfnamefont {J.}~\bibnamefont {Borregaard}},\ }\href@noop {} {\  (\bibinfo {year} {2024})},\ \Eprint {http://arxiv.org/abs/2401.15066} {arXiv:2401.15066 [quant-ph]} \BibitemShut {NoStop}%
\bibitem [{\citenamefont {Noh}\ \emph {et~al.}(2019)\citenamefont {Noh}, \citenamefont {Albert},\ and\ \citenamefont {Jiang}}]{channels}%
  \BibitemOpen
  \bibfield  {author} {\bibinfo {author} {\bibfnamefont {K.}~\bibnamefont {Noh}}, \bibinfo {author} {\bibfnamefont {V.~V.}\ \bibnamefont {Albert}}, \ and\ \bibinfo {author} {\bibfnamefont {L.}~\bibnamefont {Jiang}},\ }\href {\doibase 10.1109/TIT.2018.2873764} {\bibfield  {journal} {\bibinfo  {journal} {IEEE Transactions on Information Theory}\ }\textbf {\bibinfo {volume} {65}},\ \bibinfo {pages} {2563} (\bibinfo {year} {2019})}\BibitemShut {NoStop}%
\bibitem [{\citenamefont {Knill}(2005)}]{knillstyle}%
  \BibitemOpen
  \bibfield  {author} {\bibinfo {author} {\bibfnamefont {E.}~\bibnamefont {Knill}},\ }\href {\doibase 10.1038/nature03350} {\bibfield  {journal} {\bibinfo  {journal} {Nature}\ }\textbf {\bibinfo {volume} {434}},\ \bibinfo {pages} {39} (\bibinfo {year} {2005})}\BibitemShut {NoStop}%
\bibitem [{\citenamefont {Bernardes}\ \emph {et~al.}(2011)\citenamefont {Bernardes}, \citenamefont {Praxmeyer},\ and\ \citenamefont {van Loock}}]{nadja_rateanalysis}%
  \BibitemOpen
  \bibfield  {author} {\bibinfo {author} {\bibfnamefont {N.~K.}\ \bibnamefont {Bernardes}}, \bibinfo {author} {\bibfnamefont {L.}~\bibnamefont {Praxmeyer}}, \ and\ \bibinfo {author} {\bibfnamefont {P.}~\bibnamefont {van Loock}},\ }\href {\doibase 10.1103/PhysRevA.83.012323} {\bibfield  {journal} {\bibinfo  {journal} {Phys. Rev. A}\ }\textbf {\bibinfo {volume} {83}},\ \bibinfo {pages} {012323} (\bibinfo {year} {2011})}\BibitemShut {NoStop}%
\bibitem [{\citenamefont {Shchukin}\ \emph {et~al.}(2019)\citenamefont {Shchukin}, \citenamefont {Schmidt},\ and\ \citenamefont {van Loock}}]{waitingtime_pra}%
  \BibitemOpen
  \bibfield  {author} {\bibinfo {author} {\bibfnamefont {E.}~\bibnamefont {Shchukin}}, \bibinfo {author} {\bibfnamefont {F.}~\bibnamefont {Schmidt}}, \ and\ \bibinfo {author} {\bibfnamefont {P.}~\bibnamefont {van Loock}},\ }\href {\doibase 10.1103/PhysRevA.100.032322} {\bibfield  {journal} {\bibinfo  {journal} {Phys. Rev. A}\ }\textbf {\bibinfo {volume} {100}},\ \bibinfo {pages} {032322} (\bibinfo {year} {2019})}\BibitemShut {NoStop}%
\bibitem [{\citenamefont {Kamin}\ \emph {et~al.}(2023)\citenamefont {Kamin}, \citenamefont {Shchukin}, \citenamefont {Schmidt},\ and\ \citenamefont {van Loock}}]{rateanalysis}%
  \BibitemOpen
  \bibfield  {author} {\bibinfo {author} {\bibfnamefont {L.}~\bibnamefont {Kamin}}, \bibinfo {author} {\bibfnamefont {E.}~\bibnamefont {Shchukin}}, \bibinfo {author} {\bibfnamefont {F.}~\bibnamefont {Schmidt}}, \ and\ \bibinfo {author} {\bibfnamefont {P.}~\bibnamefont {van Loock}},\ }\href {\doibase 10.1103/PhysRevResearch.5.023086} {\bibfield  {journal} {\bibinfo  {journal} {Phys. Rev. Res.}\ }\textbf {\bibinfo {volume} {5}},\ \bibinfo {pages} {023086} (\bibinfo {year} {2023})}\BibitemShut {NoStop}%
\bibitem [{\citenamefont {Budinger}\ \emph {et~al.}(2024)\citenamefont {Budinger}, \citenamefont {Furusawa},\ and\ \citenamefont {van Loock}}]{niklas_cpg}%
  \BibitemOpen
  \bibfield  {author} {\bibinfo {author} {\bibfnamefont {N.}~\bibnamefont {Budinger}}, \bibinfo {author} {\bibfnamefont {A.}~\bibnamefont {Furusawa}}, \ and\ \bibinfo {author} {\bibfnamefont {P.}~\bibnamefont {van Loock}},\ }\href {\doibase 10.1103/PhysRevResearch.6.023332} {\bibfield  {journal} {\bibinfo  {journal} {Phys. Rev. Res.}\ }\textbf {\bibinfo {volume} {6}},\ \bibinfo {pages} {023332} (\bibinfo {year} {2024})}\BibitemShut {NoStop}%
\bibitem [{\citenamefont {Chuang}\ \emph {et~al.}(1997)\citenamefont {Chuang}, \citenamefont {Leung},\ and\ \citenamefont {Yamamoto}}]{ADcodes}%
  \BibitemOpen
  \bibfield  {author} {\bibinfo {author} {\bibfnamefont {I.~L.}\ \bibnamefont {Chuang}}, \bibinfo {author} {\bibfnamefont {D.~W.}\ \bibnamefont {Leung}}, \ and\ \bibinfo {author} {\bibfnamefont {Y.}~\bibnamefont {Yamamoto}},\ }\href {\doibase 10.1103/PhysRevA.56.1114} {\bibfield  {journal} {\bibinfo  {journal} {Phys. Rev. A}\ }\textbf {\bibinfo {volume} {56}},\ \bibinfo {pages} {1114} (\bibinfo {year} {1997})}\BibitemShut {NoStop}%
\bibitem [{\citenamefont {Bennett}\ \emph {et~al.}(1997)\citenamefont {Bennett}, \citenamefont {DiVincenzo},\ and\ \citenamefont {Smolin}}]{erasurechannel}%
  \BibitemOpen
  \bibfield  {author} {\bibinfo {author} {\bibfnamefont {C.~H.}\ \bibnamefont {Bennett}}, \bibinfo {author} {\bibfnamefont {D.~P.}\ \bibnamefont {DiVincenzo}}, \ and\ \bibinfo {author} {\bibfnamefont {J.~A.}\ \bibnamefont {Smolin}},\ }\href {\doibase 10.1103/PhysRevLett.78.3217} {\bibfield  {journal} {\bibinfo  {journal} {Phys. Rev. Lett.}\ }\textbf {\bibinfo {volume} {78}},\ \bibinfo {pages} {3217} (\bibinfo {year} {1997})}\BibitemShut {NoStop}%
\bibitem [{\citenamefont {Schumacher}\ and\ \citenamefont {Nielsen}(1996)}]{Schumacher1996}%
  \BibitemOpen
  \bibfield  {author} {\bibinfo {author} {\bibfnamefont {B.}~\bibnamefont {Schumacher}}\ and\ \bibinfo {author} {\bibfnamefont {M.~A.}\ \bibnamefont {Nielsen}},\ }\href {\doibase 10.1103/PhysRevA.54.2629} {\bibfield  {journal} {\bibinfo  {journal} {Phys. Rev. A}\ }\textbf {\bibinfo {volume} {54}},\ \bibinfo {pages} {2629} (\bibinfo {year} {1996})}\BibitemShut {NoStop}%
\bibitem [{\citenamefont {Lloyd}(1997)}]{Lloyd1997}%
  \BibitemOpen
  \bibfield  {author} {\bibinfo {author} {\bibfnamefont {S.}~\bibnamefont {Lloyd}},\ }\href {\doibase 10.1103/PhysRevA.55.1613} {\bibfield  {journal} {\bibinfo  {journal} {Phys. Rev. A}\ }\textbf {\bibinfo {volume} {55}},\ \bibinfo {pages} {1613} (\bibinfo {year} {1997})}\BibitemShut {NoStop}%
\end{thebibliography}%
\bibliographystyle{apsrev4-1}

\begin{acknowledgments}
    
We acknowledge funding from the BMBF in Germany 
(QR.N and former QR.X, PhotonQ, QuKuK, QuaPhySI), 
from the Deutsche Forschungsgemeinschaft 
(DFG, German Research Foundation) –
Project-ID 429529648 – TRR 306 QuCoLiMa (“Quantum
Cooperativity of Light and Matter”), and from the
EU project CLUSTEC (grant agreement no. 101080173).
\end{acknowledgments}

\appendix

\section{Initial state generation}
\label{app::sec::stategen}
A GKP state generation method based on controlled phase rotation between the target mode initialized in a momentum-squeezed state and a coherent-state ancilla was already proposed in the original GKP paper \cite{gkp}. The idea is to rotate the ancilla's coherent state in phase-space by an angle correlated to the target mode's position quadrature. A subsequent measurement of the rotation angle, e.g. by means of heterodyne detection, implements a modular position measurement on the target mode, collapsing the superposition to retain only such contributions where the target's position quadrature is consistent with the measured angle. The resulting state's position wavefunction is a sum of evenly spaced peaks whose width is related to the amplitude of the coherent state used as the ancilla, modulated by a wide Gaussian arising from the original momentum-squeezed state's wavefunction.  
The controlled phase rotation can be decomposed into a series of simpler gates, as detailed in Ref.~\cite{niklas_cpg}. Firstly, a so-called two-mode cubic QND gate $e^{i\alpha x_1 x_2^2} $ can be realized exactly as a series of single-mode cubic phase gates on the target (mode 1) alternating with beam-splitter type interactions between ancilla (mode 2) and target (see Eq.~(14) in Ref.~\cite{niklas_cpg}). 
Assuming the existence of an on-demand source of optical cubic phase states $\int dx\, e^{irx^3}\ket{x}_x$, the required single-mode cubic phase gates can be implemented by connecting the target mode with another ancilla prepared in such a cubic phase state via a SUM gate, followed by homodyne detection of this ancilla. 
Secondly, the desired controlled phase rotation is obtained approximately, up to a correctable displacement, by iterating several two-mode cubic QND gates, interposed with Fourier gates $F$ (i.e. phase-space rotations by $\pi/2$) on the ancilla mode. This is due to the Lie-Trotter product formula 
\begin{equation}
e^{A+B} = \lim_{k\rightarrow \infty}\left(e^{A/k}e^{B/k}\right)^k,
\end{equation}
for operators $A$ and $B$, which allows to write 
\begin{eqnarray}
&&\lim_{k\rightarrow\infty}\left(e^{i\frac{\alpha}{2k}x_1x_2^2}F_2^\dagger e^{i\frac{\alpha}{2k}x_1x_2^2}F_2\right)^k \nonumber\\
&=& \lim_{k\rightarrow\infty}\left(e^{i\frac{\alpha}{2k}x_1x_2^2} e^{i\frac{\alpha}{2k}x_1p_2^2}\right)^k \nonumber\\
&=& e^{i\alpha x_1(x_2^2 + p_2^2)/2} = e^{i\alpha x_1n_2}e^{i\alpha x_1/2}.
\end{eqnarray}

The procedure above works not only for the generation of GKP states in optical modes, but also for atomic ensemble modes, as Faraday and beam-splitter type interactions between a target ensemble and optical ancillae are possible using suitably polarized light \cite{häussler_vanloock, hammerer}. The controlled phase rotation also enables the creation of the hybrid entangled states of an ensemble and a optical multi-rail qubit forming the basis of a second generation repeater, as well between an ensemble and an optical GKP qubit, as required for the fourth generation.

\section{Statistics of memory waiting time}
\label{app::sec::statistics}
Since the entanglement distribution process in each segment of the repeater follows a geometric distribution, the number of timesteps $W$ that modes of neighboring segments have to wait for each other corresponds to the absolute value of the difference of two geometric random variables. 
Let $\mathcal{N}_1$ and $\mathcal{N}_2$ be two geometrically distributed random variables with success probability $p$. In this section, we will derive the probability distribution as well as the expectation value of  $|\mathcal{N}_1 - \mathcal{N}_2|$. \\

The difference $\mathcal{N}_1 - \mathcal{N}_2$ takes the value $k\in\mathds{Z}$ whenever $\mathcal{N}_1 = n$ and $\mathcal{N}_2 = n-k$ for any $n$ compatible with the condition $\mathcal{N}_{1, 2} \geq 1$, which must be fulfilled for $\mathcal{N}_1$ and $\mathcal{N}_2$ to be valid geometric random variables.
This condition is fulfilled whenever $n \geq \max(1, k+1)$, and thus the probability is given by
\begin{eqnarray}
\mathds{P}(\mathcal{N}_1 - \mathcal{N}_2 = k) &=  &\sum_{n = \max(1, k+1)}^\infty \mathds{P}(\mathcal{N}_1 = n)\mathds{P}(\mathcal{N}_2 = n-k) \nonumber\\
&= &\sum_{n = \max(1, k+1)}^\infty pq^{n-1}pq^{n-k-1}\nonumber\\
& = &\frac{p^2}{q^{2+k}}\sum_{n = \max(1, k+1)}^\infty (q^2)^n,
\label{app::eq::stat1}
\end{eqnarray}
where $q = 1-p$. 
The sum in Eq.~(\ref{app::eq::stat1}) is reminiscent of a geometric series; however, the lower bound is not 0 and the sum needs to be rewritten as $\sum_{n=m}^\infty = \sum_{n=0}^\infty - \sum_{n=0}^{m-1}$ before the well-known results about the limit and the finite partial sums of the geometric series can be applied. At this stage, it makes sense to introduce a case distinction between positive and strictly negative~$k$
\begin{equation}
    \mathds{P}(\mathcal{N}_1 - \mathcal{N}_2 = k) = \frac{p^2}{q^{2+k}}
    \begin{cases}
        \sum_{n = k+1}^\infty (q^2)^n & k\geq 0\\
        \sum_{n = 1}^\infty (q^2)^n & k < 0
    \end{cases}
\end{equation}
and then perform the rewrite on both cases:
\begin{equation}
\mathds{P}(\mathcal{N}_1 - \mathcal{N}_2 = k) =  \frac{p^2}{q^{2+k}} 
\begin{cases}
\left[\frac{1}{1-q^2} - \frac{1-q^{2k+2}}{1-q^2}\right] & k\geq 0 \\
\left[\frac{1}{1-q^2} - 1\right] & k < 0.
\end{cases}
\end{equation}
Simplifying, one arrives at
\begin{equation}
\mathds{P}(\mathcal{N}_1 - \mathcal{N}_2 = k) =  
\begin{cases}
\frac{p^2q^k}{1 - q^2} & k\geq 0 \\
\frac{p^2}{q^k(1 - q^2)} & k < 0,
\end{cases}
\end{equation}
from which the distribution of the difference's absolute value can be read off immediately:
\begin{equation}
\label{app::eq::probdist}
\mathds{P}(|\mathcal{N}_1 - \mathcal{N}_2| = k) = \begin{cases}\frac{p^2}{1-q^2} & k = 0\\
\frac{2p^2q^k}{1-q^2} & k > 0.\end{cases}
\end{equation}
 The expectation value $\mathds{E}(|\mathcal{N}_1 - \mathcal{N}_2|)$ is also easy to calculate:
\begin{eqnarray}
\mathds{E}(|\mathcal{N}_1 - \mathcal{N}_2|) &= &\sum_{k = 0}^\infty k \mathds{P}(|\mathcal{N}_1 - \mathcal{N}_2| = k) = \frac{2p^2}{1-q^2}\sum_{k=1}^\infty kq^k \nonumber\\
 &= &\frac{2p^2}{1-q^2}\frac{q}{(q-1)^2} = \frac{2q}{1-q^2},
\end{eqnarray}
where use was made of the identity $\sum_{k = 0}^\infty kq^k = \frac{q}{(q-1)^2}$.
For the random variable $W$, distributed according to Eq.(\ref{app::eq::probdist}), the expectation value will get denoted by $T = \mathds{E}(W)$.

\section{Amplification strategies}
\label{app::sec::amplification}
Amplification is required in order to convert bosonic losses naturally occurring in a quantum repeater into Gaussian shifts that can be corrected using the GKP code. 
A phase-insensitive amplification channel acting on a state $\rho$ is defined as a two-mode squeezing interaction with an ancillary vacuum mode:
\begin{equation}
\label{app::eq::ampchannel}
\mathcal{A}_\lambda[\rho] = \text{tr}_E \left(S_2(\lambda)\rho\otimes\ket{0}_E\bra{0}S_2^\dagger(\lambda)\right).
\end{equation}
Here, $S_2(\lambda) = e^{\text{arcosh}(\sqrt{\lambda})(a_S^\dagger a_E^\dagger - a_S a_E)}$ is the unitary operator corresponding to a two-mode squeezing interaction between the system $S$ and the environment $E$ with squeezing strength $\lambda$.
Composition of such a channel with a bosonic loss channel with matching transmissivity $\eta = 1/\lambda$ results in a Gaussian displacement channel.
If the amplification occurs before the loss, it is referred to as preamplification, otherwise as postamplification. 

Another interesting amplification method applicable in conjunction with a Bell state measurement was proposed in \cite{alloptical}. It does not require a physical interaction of the form of Eq.~(\ref{app::eq::ampchannel}) with the data mode but only a rescaling of measurement results on a classical computer; however, its use is restricted to the case of equally strong losses on both modes involved in the Bell measurement. In this section, we will discuss how to incorporate amplification into our repeater schemes and derive the variances resulting from the different strategies.  

\subsection{Amplifying memory loss }
\label{app::sec::amplification_wait}
Since entanglement distribution in each segment is a stochastic process, it may occur that a mode of a successfully distributed Bell-pair needs to wait in memory for the neighboring segment to also successfully complete distribution.
During this waiting time, the ``old'' mode undergoes a loss channel with transmissivity
\begin{equation}
\eta_\text{wait} = \exp(-\alpha W),
\end{equation}
with the dimensionless decay constant $\alpha$ related to the memory coherence time $\tcoh$ via $\alpha = \tau/\tcoh$ and $W$ representing how many timesteps of length $\tau_0$ the ``old'' mode has to wait.

It is a well-known fact that when converting a loss channel of transmissivity $\eta$ to a Gaussian shift channel, preamplification leads to a lower variance of $1 - \eta$ than postamplification with $(1 - \eta)/\eta$, and therefore should be preferred. However, preamplification is not straightforward in the case of a repeater based on entanglement swapping as soon as possible, due to the waiting time being a random variable and thus the amplitude of the loss channel not being known a priori. 
A possible solution to this problem consists in preamplifying at the beginning of each timestep with the amplification strength matching the loss during one timestep, and depending on whether the neighboring segment heralds success or not, either performing entanglement swapping or preamplification for the next timestep. 
The sum of the variances incurred by both the ``old'' and the ``new'' modes in this process, in addition to the finite squeezing of $2\delta^2$, is given by
\begin{equation}
\label{app::eq::varwait_pre}
\sigma^2_\text{add} = (W + 2)(1 - e^{-\alpha}),
\end{equation}
as both modes are kept in memory for at least one time step while the multi-rail photons propagate to the midpoint of their respective segment and the information about the success of the Bell measurement is sent back to the stations, and the ``old'' mode further receives a variance of $1 - e^{-\alpha}$ from preamplifiaction for each time step it spends waiting.  
The expectation value of Eq.~(\ref{app::eq::varwait_pre}) is obviously
\begin{equation}
    \mathds{E}(\sigma^2_\text{add}) = (T + 2)(1 - e^{-\alpha}), 
\end{equation}
with $T = 2q / (1 - q^2)$ as derived in App.~\ref{app::sec::statistics}.\\

The issue of missing information about the loss channel transmissivity could be avoided using postamplification. A better solution, however, consists in the method of ``CC-amplification'' \cite{alloptical}, 
for which it is a necessary condition that the two modes on which the Bell measurement is to be performed later be subject to losses of equal strength. In our case this means that an artificial loss channel whose amplitude corresponds to the loss experienced by the ``old'' mode must be applied to the ``new'' mode before the swapping can take place.
On atomic ensemble memories, such an artificial loss channel can be simulated by a beamsplitter-type interaction \cite{hammerer} with an ancillary optical mode.
The additional variance in this case is given by
\begin{equation}
\label{app::eq::varwait_cc}
\sigma_\text{add}^2 = \frac{1-e^{-(W+1)\alpha}}{e^{-(W+1)\alpha}}.
\end{equation}

Now we will derive the expectation value of Eq.~(\ref{app::eq::varwait_cc}) based on the distribution of the waiting time for a single segment as presented in App.~\ref{app::sec::statistics}. 
We have
\begin{eqnarray}
\mathds{E}(\sigma_\text{add}^2) &= &\frac{p^2}{1-q^2} \frac{1 - e^{-\alpha}}{e^{-\alpha}} + \sum_{N=1}^\infty \frac{2p^2}{1-q^2} q^N\frac{1-e^{-\alpha(N+1)}}{e^{-\alpha(N+1)}} \nonumber\\
&= &\frac{p^2}{1-q^2} \frac{1 - e^{-\alpha}}{e^{-\alpha}} + \sum_{N=1}^\infty \frac{2p^2}{1-q^2} q^N \left(e^{\alpha(N+1)} - 1\right) \nonumber\\
&= &\frac{p^2}{1-q^2} \left(\frac{1 - e^{-\alpha}}{e^{-\alpha}} \right.\nonumber\\
&&+  \left.2e^{\alpha}\sum_{N=1}^\infty (e^\alpha q)^N - 2\sum_{N=1}^\infty q^N\right).\quad
\end{eqnarray}
Under the assumption $q < e^{-\alpha}$ we can make use of the geometric series limit and write
\begin{equation}
\mathds{E}(\sigma_\text{add}^2) = \frac{p^2}{1-q^2} \left[\frac{1 - e^{-\alpha}}{e^{-\alpha}} + \frac{2qe^{2\alpha}}{1 - qe^\alpha} - \frac{2q}{1-q}\right].
\end{equation}\\

One observes that CC-amplification performs best as long as the entanglement distribution failure probability is not too high, i.e. the segment length $L_0$ is not too long and/or $\plink$ is sufficiently good. 
Another scheme based on preamplifying with a strength adapted to the average waiting time $T$ was investigated in Ref.~\cite{häussler_vanloock} but found never to perform better than CC-amplification or single-timestep preamplification.
When performing the rate analysis, we always assume the amplification strategy that yields the smallest variance, thus setting the expected variance to the minimum of Eqs.~(\ref{app::eq::varwait_pre}) and (\ref{app::eq::varwait_cc}), which corresponds exactly to Eq.~(\ref{eq::varwait}) in Sec.~\ref{sec::GKP}.

\subsection{Amplifying transmission loss}
\label{app::sec::amplification_dist}
The amplification to compensate for transmission loss is more straightforward than that for memory loss, since no random variables are involved. For a loss channel with transmissivity $\eta_\text{dist}$, CC-amplification leads to
\begin{equation}
    \chi^2_\text{add} = \frac{1 - \eta_\text{dist}}{\eta_\text{dist}},
\end{equation}
while preamplification leads to
\begin{equation}
    \chi^2_\text{add} = 2(1 - \eta_\text{dist}).
\end{equation}
It follows immediately that CC-amplification should be preferred whenever $\eta_\text{dist} > 1/2$. For all stronger losses, preamplificaction would result in a smaller variance; however, there is a limit to the practically achievable strength of preamplification, since it requires a physical two-mode squeezing interaction. We introduce the parameter $\eta_\text{thresh}$ as the lowest loss channel transmissivity allowing for compensation by preamplification to account for this fact.
Note that in contrast to memory loss amplification, no artificial loss channel and hence no physical interaction whatsoever is necessary to make CC-amplification work, as due to the segment's symmetric setup the states arriving from opposite directions at the midpoint are automatically subjected to equally strong loss. Therefore, even for transmissivities below $\eta_\text{thresh}$, CC-amplification is a valid option.  
Consequently, the best strategy is preamplification if $\eta_\text{thresh} \leq \eta_\text{dist} \leq 1/2$ and CC-amplification in all other cases.

\section{The fourth generation from a more general perspective}
\label{app::sec::generalconcept}
In this section we discuss how the fourth generation as a general concept fits into the framework of the established repeater generations. To answer this question, we require a figure of merit that captures the unique characteristics of each generation while retaining generality and not specifying a certain QEC code. These ``fundamental rates'' are expressed as a function of the single-segment transmissivity $\eta_0$ and defined for the various generations as follows:
For the point-to-point link (PPL) we choose, up to a constant factor, the two-way capacity of the bosonic loss channel \cite{plob} with transmissivity $\eta = \eta_0^n$ spanning the entire communication distance from Alice to Bob:
\begin{equation}
F_\text{PPL} = -1000\log_2\left(1 - \eta_0^n\right).
\end{equation}
The factor of $1000$ that will also appear for the third generation takes into account the different repetition rates of memory-based and memoryless schemes, with the former being restricted by two-way classical communication and the latter only by local state processing. 
While the impact of the two-way classical communication
for the memory-based case is in general stronger 
than a decrease by one thousandth, in the regime  
of short segment lengths approaching $L_0 = 100$m,
the elementary time units for communicating between
neighboring stations match the typical local processing
times of the light-matter interfaces, $\text{MHz}^{-1}$.
In contrast, a memoryless, all-optical scheme is
primarily limited by the elementary source and 
detector times which are of the order $\text{GHz}^{-1}$.
For the second generation we simply use the loss channel's repeater-assisted capacity \cite{repeaterassisted}
\begin{equation}
\label{app::eq::f2}
F_\text{2nd} = -\log_2\left(1-\eta_0\right).
\end{equation}

The situation is somewhat more complex for the third and fourth generations, since we need to incorporate the effect of error correction in the channel. To this end, we make use of a ``logical'' transmissivity $\eta_0^\prime$ as an attempt to quantify how well a given QEC code protects the transmitted states from loss errors. As a reference we set $\eta_0^\prime = \eta_0$ for dual-rail single-photon transmission, and further require $\eta_0^\prime$ to equal the probability of successfully mapping the corrupted states back into the code space for loss codes, such that e.g. for the two-mode four-photon code \cite{ADcodes}
\begin{eqnarray}
    \ket{\overline{0}} &=&  \sqnorm\left(\ket{4}\ket{0} + \ket{0}\ket{4}\right)\nonumber\\
    \ket{\overline{1}} &=& \ket{2}\ket{2},
\end{eqnarray}
where the kets on the right-hand side denote Fock states of the respective modes and where two physical amplitude damping channels with equal transmissivities $\eta_0$ act independently and individually on the two modes,
we get 
\begin{equation}
    \eta_0^\prime = \eta_0^4 + 4\eta_0^3(1 - \eta_0).
\end{equation}
For the GKP code, where unheralded Pauli errors come into play, we propose to define the ``logical'' transmissivity as 
\begin{equation}
\label{app::eq::gkptransmissivity}
    \eta_0^\prime = p(1 - 2H(\ppauli{dist})),
\end{equation}
with $p$ and $\ppauli{dist}$ given by Eqs.~(\ref{eq::psucc4}) and (\ref{eq::ppaulidist}), respectively. This is motivated by the following consideration: We are interested in finding such a transmissivity $\eta_0^\prime$ that the code in question performs equally well under the physical $\eta_0$ as a dual-rail photon under a fictitious loss channel with $\eta_0^\prime$ (in the two-mode case again corresponding to two independent and individual amplitude damping channels with equal transmissivities, now replacing $\eta_0$ by $\eta_0^\prime$), where the logical information in a dual-rail qubit experiences an erasure channel with erasure probability $p_e = 1 - \eta_0^\prime$. The (two-way) capacity of such a channel is known \cite{erasurechannel} to be $1 - p_e = \eta_0^\prime$, while the capacity of a Pauli-channel is lower-bounded \cite{plob} by $1 - 2H(\ppauli{})$. Setting these expressions equal, and additionally accounting for a possibly non-unit HRM acceptance probability, results in Eq.~(\ref{app::eq::gkptransmissivity}). 

Given the ``logical'' transmissivity, we can formulate the third generation's ``fundamental rate'' by pretending we were sending dual-rail qubits along segments characterized by $\eta_0^\prime$. The individual transmissivities behave multiplicatively to form the total transmssivity $(\eta_0^\prime)^n$, such that we obtain
\begin{equation}
\label{app::eq::f3}
F_\text{3rd} =  -1000\log_2\left(1 - (\eta_0^\prime)^n\right),
\end{equation}
again including a factor of $1000$, as for the PPL. Note that in general, $(\eta_0^\prime)^n$ might not equal $\eta^\prime$, i.e. multiplication of transmissivities does not necessarily commute with the ``priming''-operation.
Further note that while Eq.~(\ref{app::eq::f3}) does not explicitly account for the one-way nature of classical communication between any two neighboring repeater stations in a third-generation protocol, we still believe it to be a more meaningful expression for capturing the unique characteristics of the third generation than e.g. the one-way capacity of a single segment, since the latter does not exhibit the typical unfavorable $n$-scaling. For a fixed total distance, one could thus shorten the individual segments at will without punishment, eventually approaching the two-way capacity in each segment and effacing any distinction with a memory-based scheme.  
While our expression is ostensibly plagued by the shortcoming of not explicitly enforcing that a single-segment transmissivity lower than $1/2$ result in a vanishing rate, this will not impact the discussion in the following, because in such a case the ``logical'' transmissivity cannot surpass the physical transmissivity, and as we will see shortly, this ensures that the third generation will not be the optimal one. 

The fourth generation can be approached in a similar way: When abstracting from the physical reality of the state transmission and pretending we were working with dual-rail photons in an $\eta_0^\prime$-channel, the second and fourth generations become the same protocol, and therefore we simply set
\begin{equation}
F_\text{4th} = -\log_2\left(1 - \eta_0^\prime\right)
\end{equation}
in analogy to Eq.~(\ref{app::eq::f2}).
Of course, in reality, all schemes are still bounded by the repeater-assisted bound \cite{repeaterassisted} of $-\log_2(1 - \eta_0)$ with the physical transmissivity $\eta_0$; however, the ``fundamental rates'' are a helpful tool for comparing schemes. For example, since second and fourth generation are essentially the same protocol up to the details of entanglement distribution, we might assume that for a given QEC code, the rate of the fourth generation saturates the ``false'' bound $F_\text{4th}$ to a similar degree as that of the second generation saturates the ``real'' bound $F_\text{2nd}$, and thus investigating when $F_\text{4th}$ can surpass $F_\text{2nd}$ may serve as an estimation of when a given instance of the fourth generation can beat that of the second. \\

\begin{figure*}
\begin{center}
\includegraphics[width=\linewidth]{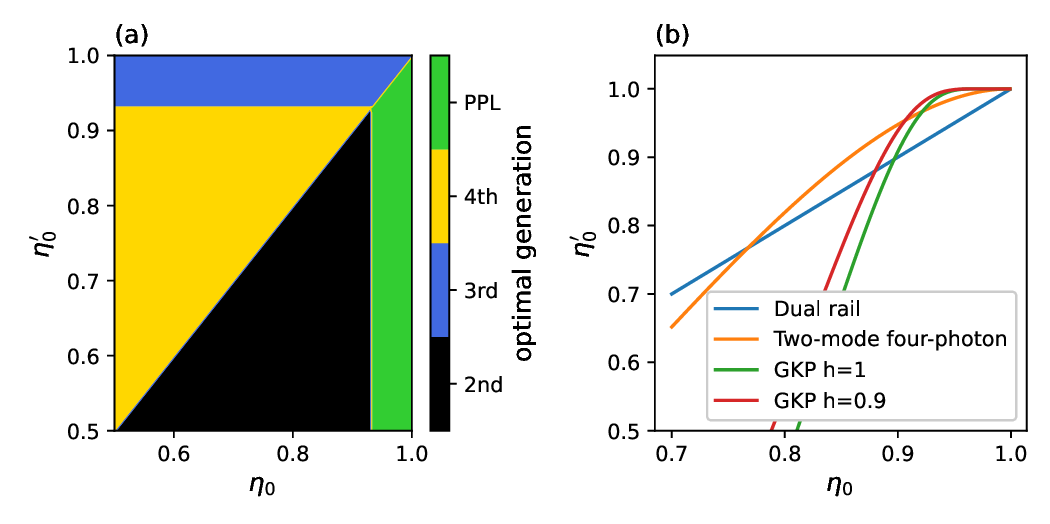}
\end{center}
\caption{(Color online) (a) The $\eta_0$-$\eta_0^\prime$ plane is divided into areas where a given repeater generation is expected to have an advantage over the others according to Eqs.~(\ref{app::eq::conditionsstart}) to (\ref{app::eq::conditionsend}) for $n=100$. With more segments, the intersection point at which all areas meet shifts along the diagonal towards the upper right. (b) ``Logical'' transmissivity $\eta_0^\prime$ as a function of the physical transmissivity $\eta_0$ for various codes. The curve for the dual-rail photon corresponds to the diagonal $\eta_0^\prime = \eta_0$. The GKP curves are for qubits in the infinite-squeezing limit $s\rightarrow \infty$.}
\label{fig::generalconcept}
\end{figure*}

While for any specific code $\eta_0^\prime$ can be expressed as a function of $\eta_0$, we will keep the former as a free parameter and study how the generations' performances relate to each other for any point in an $\eta_0$-$\eta_0^\prime$ plane. 
It is immediately apparent that we have $F_\text{3rd} > F_\text{PPL}$ and $F_\text{4th} > F_\text{2nd}$ whenever $\eta_0^\prime > \eta_0$.
For PPL vs. fourth generation, we find
\begin{eqnarray}
\label{app::eq::conditionsstart}
\eta_0^\prime & > 1 - \left(1 - \eta_0^n\right)^{1000} \qquad & \text{4th better}\nonumber\\
\eta_0^\prime & < 1 - \left(1 - \eta_0^n\right)^{1000}  \qquad & \text{PPL better},
\end{eqnarray}
and similarly we obtain 
\begin{eqnarray}
\eta_0^\prime & > \left[1 - (1 - \eta_0)^{1/1000}\right]^{1/n} \qquad & \text{3rd better}\nonumber\\
\eta_0^\prime & < \left[1 - (1 - \eta_0)^{1/1000}\right]^{1/n}  \qquad & \text{2nd better}
\end{eqnarray}
for second generation vs. third generation.
To find the intersections of $F_\text{PPL}$ with $F_\text{2nd}$ and $F_\text{3rd}$ with $F_\text{4th}$, we make use of the 
approximation $(1-x)^{1000} \approx 1 - 1000x$, valid for small $x$, and reduce the problem to solving $1 - \eta_0 = 1 - 1000\eta_0^n$ (and similarly with $\eta_0^\prime$) to obtain:
\begin{eqnarray}
\eta_0 & > (1/1000)^{1/(n-1)} \qquad & \text{PPL better}\nonumber\\
\eta_0 & < (1/1000)^{1/(n-1)}\qquad & \text{2nd better}
\end{eqnarray}
as well as
\begin{eqnarray}
\label{app::eq::conditionsend}
\eta_0^\prime & > (1/1000)^{1/(n-1)} \qquad & \text{3rd better}\nonumber\\
\eta_0^\prime & < (1/1000)^{1/(n-1)}  \qquad & \text{4th better}.
\end{eqnarray}
The areas in the $\eta_0$-$\eta_0^\prime$ plane corresponding to these conditions are plotted in Fig.~\ref{fig::generalconcept} for the case of $100$ segments, together with the ``logical'' transmissivities as a function of the physical transmissivity for dual-rail and two-mode four-photon encoding, as well as for the GKP code with deterministic and non-deterministic distribution.  
The majority of the plane is covered by the second and fourth generations, with third and PPL occupying a stripe at the edge corresponding to very high $\eta_0^\prime$ and $\eta_0$, respectively. With a growing number of segments, the intersection point where all areas meet shifts towards the upper-right corner, further reducing the areas of the memoryless schemes, i.e. PPL and third generation. While the fourth generation seems to occupy quite a large area, it must be kept in mind that error correction codes typically are only useful if loss does not exceed a certain threshold and thus the $\eta_0^\prime$-curve only crosses above the diagonal to the right of a certain value of $\eta_0$, rendering most of the area inaccessible in practice.  
As an example, follow the GKP curves in Fig.~\ref{fig::generalconcept}(b) from low to high $\eta_0$: Initially, they lie below the diagonal and thus fall into the second-generation area. Only at about $\eta_0 = 0.88$, error correction in the channel becomes viable and the curves cross the diagonal into the fourth-generation area. After shortly passing through its upper-right corner, they reach the third-generation area when $\eta_0^\prime$ becomes so good that the advantage due to the factor $1000$ is no longer counteracted by the detrimental $n$-scaling. This is in good agreement with our findings, concerning e.g. the $\plink$-dependence, in the main text. 
Finally, especially in the context of our discussion above 
concerning the missing one-way nature of 
nearest-neighbor communication in a third-generation
protocol when represented by Eq.~(\ref{app::eq::f3}),
we note that Fig.~\ref{fig::generalconcept}(a) looks entirely unchanged
if we replace the current two-way capacities 
for the third-generation repeater and the PPL
by the corresponding (unassisted) one-way bosonic loss channel
quantum capacities \cite{plob, Schumacher1996, Lloyd1997} for the total distance, i.e.,
with transmissivities $(\eta_0^\prime)^n$ and  $\eta_0^n$, respectively,
albeit for a smaller number of segments,
replacing the current $n=100$ by $n=10$.

\end{document}